\documentclass[sigplan,screen]{acmart}

\usepackage{amsmath}
\usepackage{amsfonts}
\usepackage{multirow}
\usepackage{xspace}
\usepackage{enumitem}
\usepackage[caption=false]{subfig}
\usepackage{xurl}
\usepackage{cleveref}

\acmYear{2026}\copyrightyear{2026}
\setcopyright{cc}
\setcctype[4.0]{by}
\acmConference[SOSP '26]{Symposium on Operating Systems Principles}{September 29--October 2, 2026}{Prague, Czech Republic}
\acmBooktitle{Symposium on Operating Systems Principles (SOSP '26), September 29--October 2, 2026, Prague, Czech Republic}
\acmDOI{10.1145/3830418.3843860}
\acmISBN{979-8-4007-2585-2/26/09}

\newcommand{\sys}{\textsf{llmovoice}\xspace}
\newcommand{\Sys}{Llmovoice\xspace}
\crefformat{section}{\S#2#1#3}
\crefformat{subsection}{\S#2#1#3}
\crefformat{subsubsection}{\S#2#1#3}

\newcommand{\eg}{{\it e.g.,}\xspace}
\newcommand{\ie}{{\it i.e.,}\xspace}

\newcommand\paragraphb[1]{\noindent\textbf{#1}}
\newcommand\pb[1]{\paragraphb{#1}}

\makeatletter
\renewcommand{\@mkauthors}{%
  \global\setbox\mktitle@bx=\vbox{%
    \noindent\unvbox\mktitle@bx\par\medskip
    \centering
    {\@authorfont
      Linyi Jiang\textsuperscript{1,2}\qquad
      Silvery D. Fu\textsuperscript{2}\qquad
      Yifei Zhu\textsuperscript{1}\par}
    \vspace{0.35em}
    {\@affiliationfont
      \textsuperscript{1}Global College, Shanghai Jiao Tong University\qquad
      \textsuperscript{2}AgenticSys\par}
    \bigskip}}
\makeatother

\begin{document}

\title{Scalable Context Orchestration for Serving \\LLMs Over Voice}

\author{Linyi Jiang}
\affiliation{%
  \institution{Global College, Shanghai Jiao Tong University}
}
\affiliation{%
  \institution{AgenticSys}
}

\author{Silvery D. Fu}
\affiliation{%
  \institution{AgenticSys}
}

\author{Yifei Zhu}
\affiliation{%
  \institution{Global College, Shanghai Jiao Tong University}
}

\renewcommand{\shortauthors}{Jiang et al.}
\begin{abstract}
Voice AI applications are gaining popularity as advances in large language models (LLMs) enable more natural and accessible spoken interactions. Serving these applications requires accounting not only for what users say, but also for how they speak (\eg speaking rate) and the conditions under which their audio is captured and transmitted (\eg background noise and packet loss). However, existing LLM systems represent conversation context as a flat, growing sequence of messages, leaving voice-specific context implicit in the audio. As a result, they can generate responses that are poorly aligned with user preferences, degrade interaction quality under adverse environmental conditions, and incur high costs over long voice sessions. 

We present \sys, a context-management middleware that explicitly models voice context and orchestrates its use. At each turn, \sys constructs a bounded voice context from the current user input, relevant interaction history, and explicit paralinguistic and environmental states. It then uses the serving LLM to reason over this context and generate runtime directives that guide how the system responds. We evaluate \sys on real-world voice applications and benchmarks. It reduces speaking-rate alignment error by 52.4\%, lowers the false-interruption rate from 46.0\% to 0.9\% under packet loss, and reduces model usage cost by 79.2\%. For long sessions, \sys reduces per-turn cost by up to 24.9$\times$ while retaining up to 98.7\% of baseline answer quality.

\end{abstract}

\ccsdesc[500]{Computer systems organization}
\ccsdesc[300]{Computing methodologies~Artificial intelligence}
\ccsdesc[300]{Information systems~Information systems applications}

\keywords{voice AI, large language model applications, context management, agentic systems}

\maketitle
\raggedbottom

\pagestyle{empty}

\section{Introduction}

Voice has emerged as an important interface for AI/LLM applications~\cite{RealtimeAPIOpenAI,GeminiLiveAPI,fang2024llama,cui2025recent}. Compared with text interfaces, voice enables more natural and accessible interaction. For example, voice can extend LLM services via telephone networks to users without reliable broadband~\cite{StateMobileInternetb,gubbala3InternetSmartphone2022,khan2024artificial} and lower barriers for those with limited literacy~\cite{medhi2011designing,VoiceFirstGenerativeAI}. These advantages are driving the adoption of voice AI applications across many domains, including healthcare~\cite{chen2026speechmedassist}, education~\cite{chang2026vrcoaching}, agriculture~\cite{high2026artificial}, and in-vehicle assistance~\cite{monk2026invehicle}. 

In real-time \emph{LLM-over-voice} systems, user interaction follows a turn-by-turn streaming loop: the client application streams audio to the LLM service; once the end of the user's turn is detected, the service invokes the model and streams the generated response to the client for playback~\cite{RealtimeAPIOpenAI,GeminiLiveAPI}. Each turn, the model processes inputs in a \emph{context window} containing system prompts, the current user input, and prior conversation history. Managing context in voice interaction is challenging. First, voice interaction is often UI-less: users cannot rely on visual artifacts such as tabs to organize conversation history or visual controls to adjust how the system responds. This requires the system itself to manage the context on the user's behalf. Second, the system must account not only for \emph{what} the user says, but also for \emph{how} the user speaks and the \emph{conditions} under which the audio is captured and transmitted. In this paper, we use \emph{voice context} to refer collectively to these three types of information: (i) semantic content (\eg conversation history), (ii) paralinguistic cues (\eg speaking rate and tone), and (iii) environmental conditions, including acoustic (\eg background noise) and network conditions (\eg jitter and packet loss). 

Current LLM context-management techniques focus primarily on semantic content in text-based interactions~\cite{packer2023memgpt,lewis2020retrieval,jiang2023active} but offer limited support for the latter two forms of voice context. In particular, effective voice interaction relies on paralinguistic cues such as speaking rate, tone, and turn-taking dynamics~\cite{nass2005wired,borrie2019syncing}. For instance, a voice agent often needs to adjust its pacing (\eg faster for simple confirmations and slower for complex instructions) and adapt its delivery style to match user preferences. Existing systems leave these cues implicit in the audio, making it difficult to track them explicitly across turns and use them for runtime adaptation.
Consequently, a response can be semantically correct while its delivery remains misaligned with the user's preferences. 

Voice interaction is also highly sensitive to environmental conditions. Background noise can be mistaken for user speech, while network fluctuations such as jitter and packet loss can introduce apparent silence into the received audio. For example, because turn detection relies on observed silence, the system may mistake packet loss for the end of a turn and interrupt the user prematurely. Such false interruptions not only degrade interaction quality but also pollute the context processed in future turns, thereby increasing model usage costs.

The key challenge in managing voice context is that paralinguistic cues and environmental conditions cannot be interpreted independently of semantic content. For example, a user may explicitly ask the system to speak faster even when the user's speaking rate suggests a preference for slower delivery. Similarly, whether observed silence marks the end of a turn depends on both the semantic completeness of the current utterance and whether the silence results from network fluctuations. LLM systems must therefore explicitly maintain and jointly reason over semantic, paralinguistic, and environmental states when generating a response. Since a user instruction or preference expressed earlier may determine how the system should respond to current paralinguistic cues, the voice context required for such joint reasoning may span multiple historical turns during a voice session. 

Moreover, as conversational topics drift, branch, and recur, the system must also determine which prior information to retain in the context window. Existing systems commonly treat a voice session as a single, growing sequence of audio tokens and append as much of the conversation history as fits within the context window and reprocess it at every turn~\cite{RealtimeAPIOpenAI,GeminiLiveAPI}. This approach, however, is ineffective at selecting relevant context and is also costly: one minute of voice interaction is approximately 49$\times$ as expensive to process as the corresponding text interaction~\cite{RealtimeAPIOpenAI}, and repeatedly processing the growing history causes cumulative cost to increase quadratically with session length. Together, these challenges make high-quality, long-running voice interaction difficult to scale.

We argue that addressing these challenges is not merely a matter of implementation but one of \emph{architecture}: LLM-over-voice systems require an explicit \emph{voice-context management} layer between applications and LLM services. The role of this layer is to transform unstructured and potentially unreliable streaming voice input, together with conversation history and environmental signals, into structured voice context that can be explicitly managed across turns and used for runtime adaptation. Our key insight is that fulfilling this role requires two coupled tasks per turn: (i) constructing a \emph{bounded voice context} from the current input, history, and signals; and (ii) jointly reasoning over its semantic, paralinguistic, and environmental states to determine \emph{runtime controls}. 

In this paper, we present \sys, a voice-context management layer built around three design principles to support these tasks: (1) explicitly modeling paralinguistic and environmental states in voice context while representing semantic content at multiple levels of fidelity; (2) organizing voice context with a unified abstraction that bridges streaming voice interaction and memory; and (3) maintaining voice context through an orchestration loop that reasons jointly over its semantic, paralinguistic, and environmental states at each turn. Specifically, \sys derives paralinguistic and environmental states from incoming audio and runtime signals and represents them alongside semantic content in the voice context. Across turns, \sys materializes each completed interaction as a VoicePage (\texttt{vpage}) and groups related pages into VoiceThreads (\texttt{vthreads}) that capture higher-level conversational flows. Each \texttt{vpage} maintains its available audio, transcript, and summary representations, while each \texttt{vthread} maintains a compact summary and references to its constituent pages. This organization enables selective retrieval while preserving logical relationships among retrieved interactions.

\begin{figure}[tbp]
    \centering
    \includegraphics[width=\columnwidth]{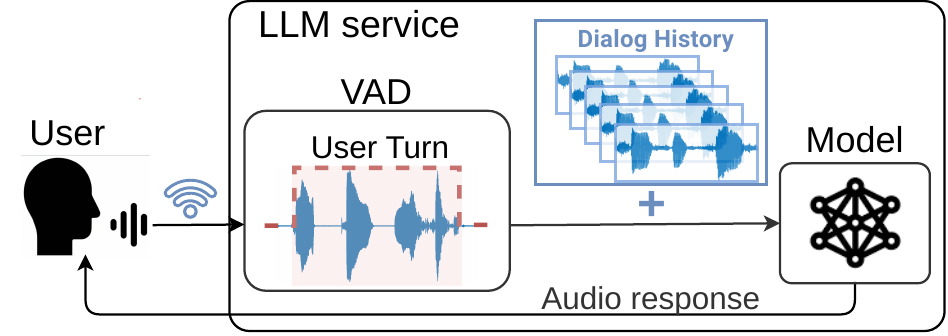}
    {\color{red}
    \vspace{-0.43em}
    \caption{Overview of the LLM-over-voice serving pipeline.
    }
\label{fig:lov_current_pipeline}
    }
\end{figure}

Upon receiving new user input, \sys retrieves relevant \texttt{vthreads} and \texttt{vpages}; a cost-aware \emph{context projector} selects a representation for each retrieved item under the available context budget. The projected history is combined with the user's current input to form the voice context, which also includes paralinguistic and environmental states. A \emph{context orchestrator} uses the serving LLM to reason jointly over these states and generate runtime directives that guide response content, adjust delivery style, and control turn-taking behavior. The orchestrator then applies these directives and calls the serving LLM to generate a response when needed. Finally, it commits the completed interaction as a new \texttt{vpage} and incorporates it into the \texttt{vthread} structure, closing the orchestration loop. Together, these abstractions allow \sys to maintain a bounded voice context that combines paralinguistic and environmental states with relevant semantic content represented at an appropriate fidelity, and to adapt runtime behavior at each turn.

We implement \sys as middleware that exposes an interface compatible with existing LLM service APIs (\eg OpenAI Realtime), enabling drop-in integration into voice applications. We evaluate \sys on real-world applications and benchmarks. Our results show that \sys improves both interaction quality and long-session scalability: it reduces speaking-rate alignment error by 52.4\%, lowers the false-interruption rate from 46.0\% to 0.9\% under packet loss, and decreases network-induced cost by 79.2\%. For long conversations, \sys reduces per-turn cost by up to 24.9$\times$ while retaining up to 98.7\% of baseline answer quality.

\section{Background and Motivation}

In this section, we provide background on existing LLM-over-voice serving pipelines and outline the key limitations that motivate our design.

\subsection{LLM-over-Voice Serving Pipeline}

LLM services increasingly support real-time, streaming voice interaction. In such systems, user audio is continuously transmitted over a persistent connection, processed incrementally, and used to generate low-latency spoken responses.

Earlier production systems typically adopt a cascaded architecture: incoming audio is first converted to text via a streaming speech-to-text (STT) module, the transcript is processed by a text-based LLM, and the output is rendered back to speech using a text-to-speech (TTS) module~\cite{huang2024audiogpt,shen2023hugginggpt}. More recent systems move toward end-to-end speech-to-speech models, where audio streams are processed directly without explicit intermediate text representations~\cite{fang2024llama,RealtimeAPIOpenAI,GeminiLiveAPI}. Compared with cascaded architectures, these speech-native models avoid separate STT and TTS stages, reducing interaction latency and preserving information that would otherwise be lost in intermediate text representations~\cite{cui2025recent}.

Specifically, interaction proceeds in a loop as illustrated in Figure~\ref{fig:lov_current_pipeline}. A client sends audio chunks over a communication channel that may drop packets. A turn-detection module, commonly voice activity detection (VAD), observes the arriving stream, determines when the user stops speaking, and commits the user turn. 
The service then invokes the model with a context window containing system prompts%
, the committed utterance as the current input, and prior conversation history, and streams the generated audio back to the client.

\subsection{Limitations of Existing Systems}

Current LLM-over-voice systems have three key limitations.

\begin{figure}[tbp]
    \centering
    \begin{minipage}[t]{0.485\columnwidth}
        \centering
        \includegraphics[width=\linewidth]{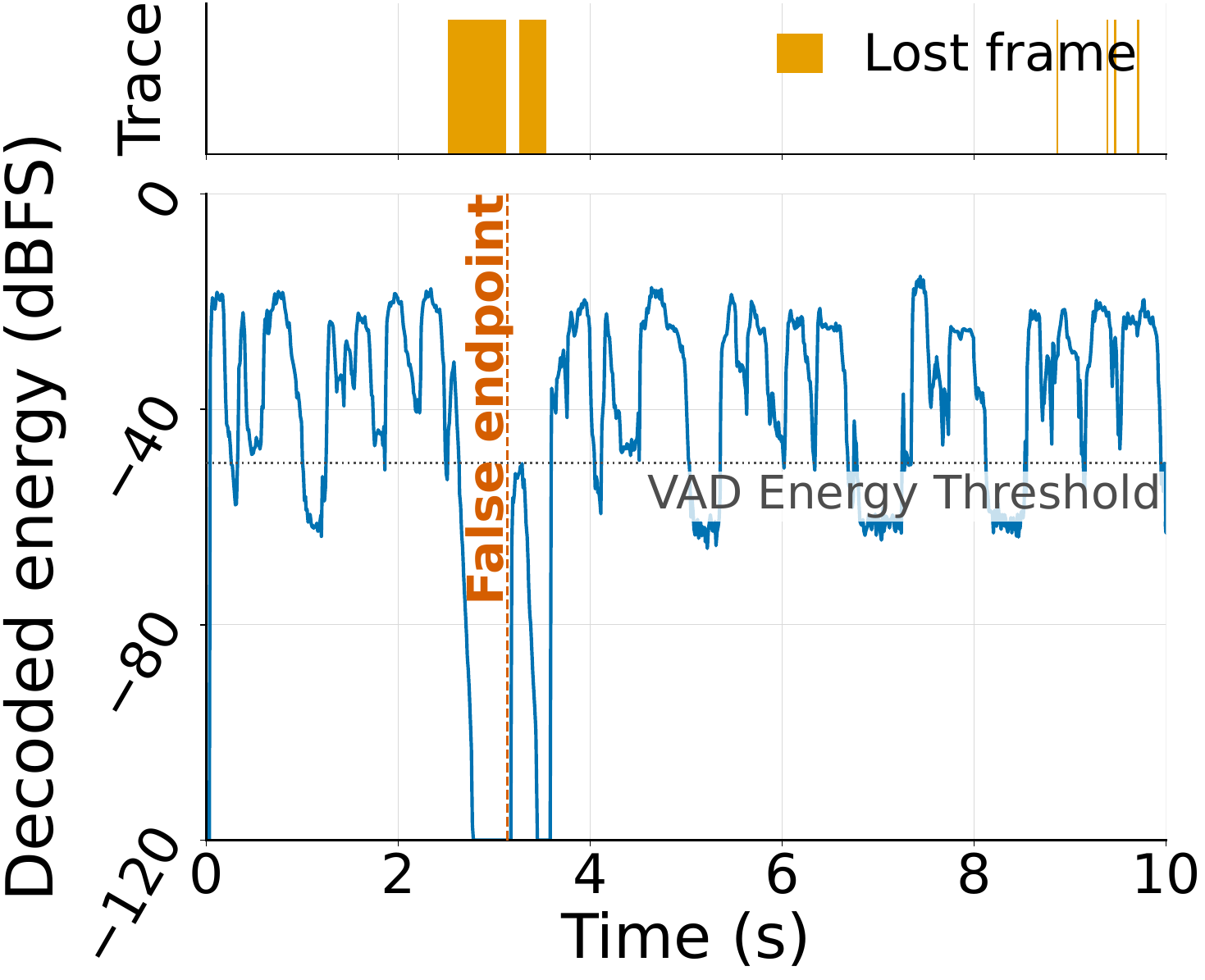}
        \captionof{figure}{Packet loss triggers a false VAD endpoint.}
        \label{fig:packet-loss-trace}
    \end{minipage}
    \hfill
    \begin{minipage}[t]{0.485\columnwidth}
        \centering
        \includegraphics[width=\linewidth]{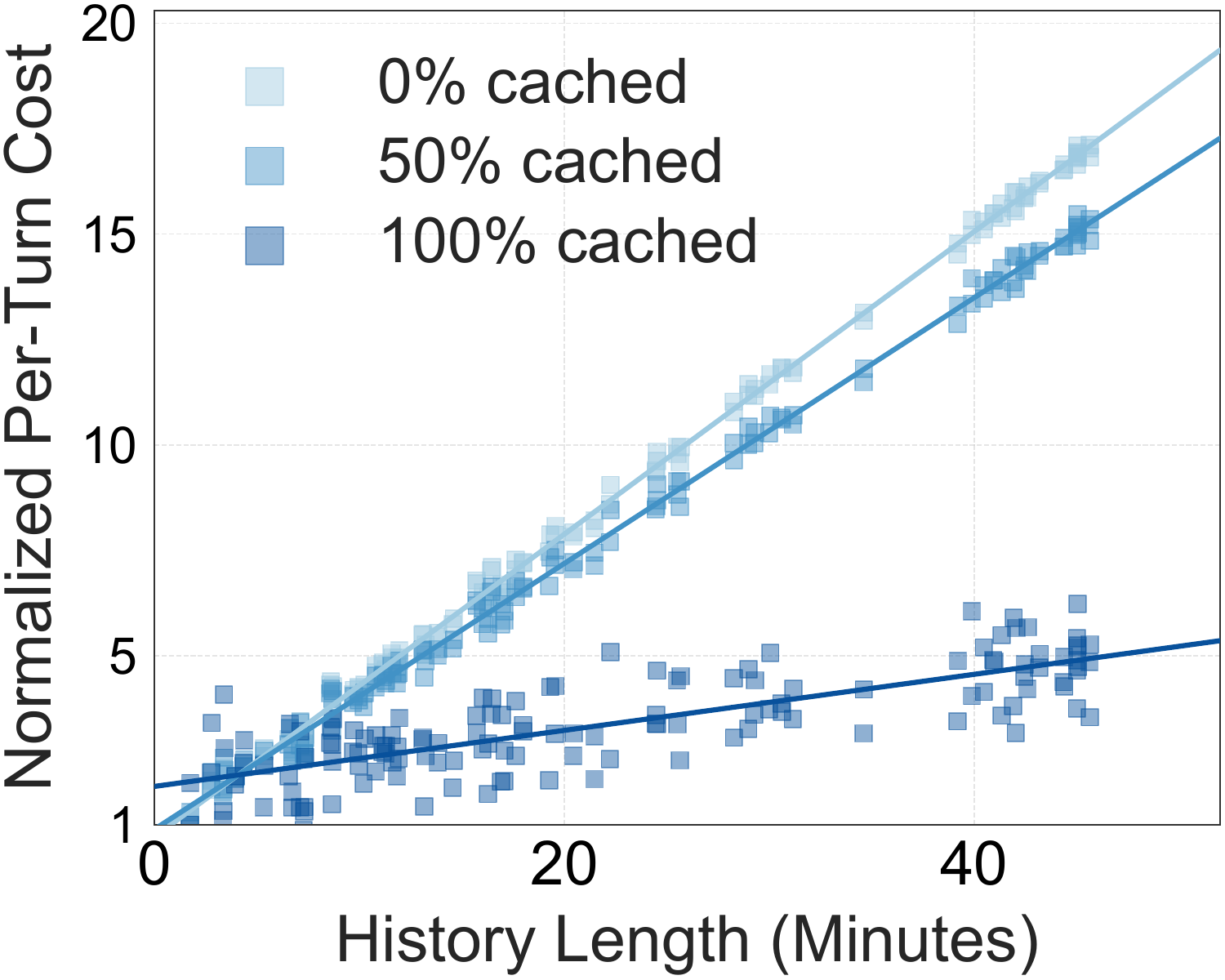}
        \captionof{figure}{Per-turn voice serving cost grows with history.}
        \label{fig:prompt-cache-scalability}
    \end{minipage}
\end{figure}

\pb{(1) Voice-specific context remains implicit.}
Existing context-management mechanisms explicitly manage semantic content but leave paralinguistic cues and environmental conditions implicit in the audio stream. Without explicit paralinguistic states, the system cannot reliably adapt its response style to the user. Conversational partners naturally align their speaking behavior, a phenomenon known as conversational entrainment~\cite{borrie2019syncing,brennan1996lexical,garrod1987saying}. For instance, a reply that is much slower than the user’s speech can feel sluggish, disrupt the conversational rhythm, and reduce user engagement. Without explicit environmental states, the system cannot distinguish natural silence from low-energy audio caused by packet loss. We illustrate this failure using a real-world packet-loss trace from the ICASSP 2024 Audio Deep PLC Challenge \cite{diener2025icassp}. Specifically, in this pilot study, we replay the validation trace \texttt{val\_0332} through a local WebRTC connection in Chromium using its default Opus pipeline. We configure the VAD proxy with representative settings: an energy threshold of $-50$\,dBFS and a silence threshold of 500\,ms. As shown in Figure~\ref{fig:packet-loss-trace}, a loss burst drives the decoded audio below the energy threshold long enough to trigger a false endpoint and prematurely invoke the model. The resulting incomplete utterance and unnecessary response can pollute the interaction history, increase model usage cost, and degrade subsequent interaction quality.

\pb{(2) State effects are not coordinated.}
Different states can imply conflicting runtime actions. Consider a user who says, ``Please respond as soon as I finish speaking,'' over a connection experiencing packet loss. The semantic state favors decreasing the VAD silence threshold to reduce response latency, while the environmental state favors increasing the same threshold to prevent packet-loss-concealed audio from being mistaken for the end of the turn. The two states therefore demand opposing updates to the same runtime parameter. Similar conflicts arise between semantic and paralinguistic states: the observed speaking rate may suggest a slower reply, while the user explicitly asks the agent to speak faster. These signals must therefore be coordinated before the runtime changes turn detection or response pacing. Separate rule-based controllers can make inconsistent decisions because each considers only one state, while a fixed ruleset cannot cover the combinations of states that arise across users, applications, and conversational turns. 

\pb{(3) Long-session context management does not scale.}
Systems that directly invoke real-time LLM APIs for voice interaction typically construct the context window by appending the conversation history as a monolithic, continuously growing sequence. Existing voice-serving systems may also support prompt caching, which reuses computation for an unchanged prefix and reduces repeated processing cost. We measure the OpenAI Realtime API on MSP-PODCAST under 0\%, 50\%, and an idealized 100\% prompt-cache hit rate (detailed experimental setup in \S\ref{sec:eval-setup}). For each cache setting, costs are normalized to the shortest session, whose cost is set to 1.0.
Figure~\ref{fig:prompt-cache-scalability} shows that per-turn cost grows approximately linearly with conversation history under different cache settings. %
This shows that caching reduces cost but not its asymptotic growth rate: per-turn cost remains linear in conversation length, yielding quadratic total session cost.
Caching also does not increase context capacity, and long-running sessions may reach the context limit after approximately 15-60 minutes~\cite{RealtimeAPIOpenAI,GeminiLiveAPI}.

\section{System Design}

To address the above limitations, \sys explicitly represents semantic, paralinguistic, and environmental states, jointly reasons over them to derive runtime controls, and replaces a monolithic full-history context with structured interaction units that can be selectively included at different fidelities. We present these voice-context management designs in this section.

\subsection{Architecture Overview}
\label{sec:architecture-overview}

\begin{figure}[t]
    \centering
    {\includegraphics[width=\columnwidth]{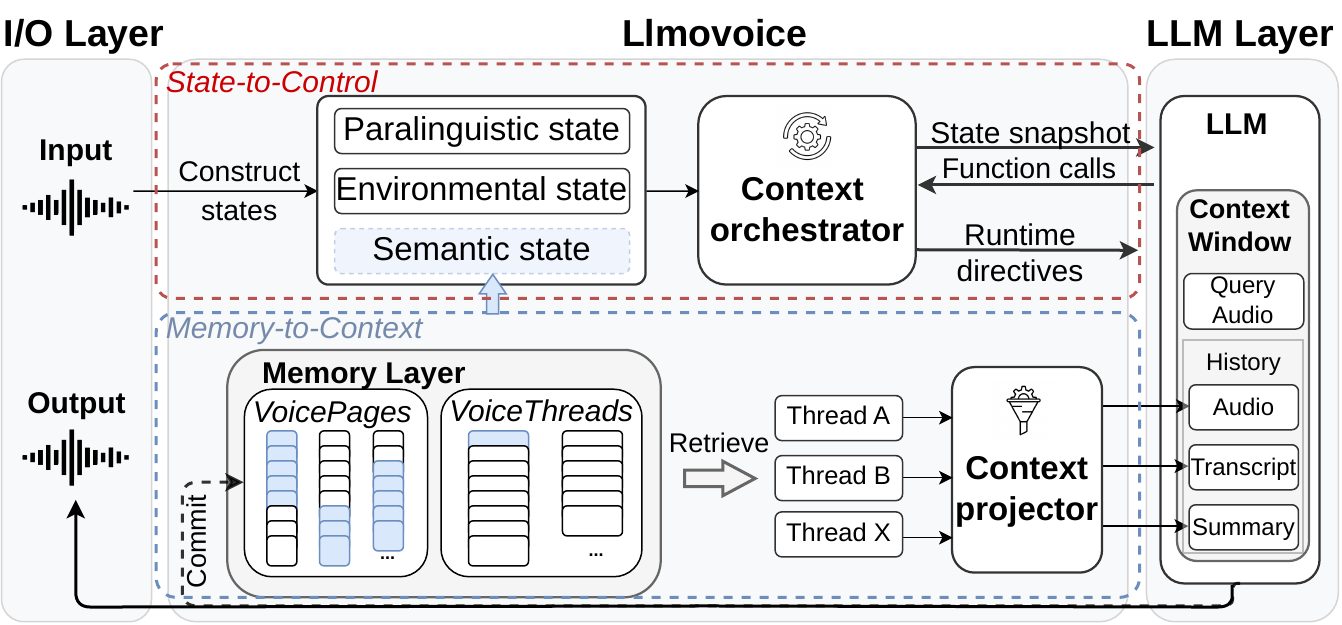}}
    \caption{System overview of \sys.}
    \label{fig:systemoverview}
\end{figure}

Figure~\ref{fig:systemoverview} depicts the architecture of \sys, which operates between an application's streaming audio interface and an LLM service. \Sys runs a voice-context orchestration loop comprising two paths: a memory-to-context path and a state-to-control path. Along the \emph{memory-to-context path}, \sys represents each completed interaction as a \texttt{vpage} containing the user input, model output, and their available representations. It organizes semantically related pages into \texttt{vthreads} that capture coherent conversational flows (\S\ref{sec:structured-history}). When a new request arrives, \sys retrieves the relevant \texttt{vthreads} and \texttt{vpages}, and the context projector selects an audio, transcript, or summary representation for each retrieved item under the available context budget (\S\ref{sec:context-projection}). The selected representations form the projected history, which is included in the context window and later used by the state-to-control path.

Along the \emph{state-to-control path}, \sys combines the projected history with the current input to form the voice context for the current turn. The
semantic state captures the semantic content of the current input together with relevant prior information from the projected history. The paralinguistic state captures how the current input is delivered, while the environmental state captures conditions that may affect how it is observed and interpreted (\S\ref{sec:state-modeling}). The context orchestrator uses the serving LLM to reason jointly over these states and generate structured function calls. It then validates and converts the resulting function calls into runtime directives that guide response content, adjust delivery style, and control
turn-taking behavior (\S\ref{sec:context-orchestration}).

After applying the directives, the serving LLM generates a response. Once the interaction completes, \sys commits the completed interaction as a new \texttt{vpage} and incorporates it into the \texttt{vthread} structure. This commit makes the interaction available to the memory-to-context path in subsequent turns, closing the voice-context orchestration loop.

\subsection{Voice Context Modeling}
\label{sec:state-modeling}

The state-to-control path extracts runtime observations and organizes them into three complementary states of voice interaction. The \textit{semantic state} captures what the interaction is about, the \textit{paralinguistic state} captures how the current input is delivered, and the \textit{environmental state} captures the acoustic and network conditions under which it is observed.

\pb{Semantic state.}
The semantic state captures the model-visible information needed to interpret the current interaction. It combines the current user input with relevant prior information from history projected by context projection (\S\ref{sec:context-projection}) in the memory-to-context path. 

\pb{Paralinguistic state.}
The paralinguistic state captures properties of speech delivery that may affect how the system responds. In this paper, we focus on speaking rate as a representative paralinguistic signal. A single short or atypical utterance may provide unreliable evidence of the user's current speaking behavior. Therefore, \sys maintains a bounded history of valid speaking-rate observations and gives greater weight to recent observations. This smooths turn-level variation while allowing the state to follow sustained changes in how the user speaks.

\pb{Environmental state.}
The environmental state captures conditions under which speech is captured or transmitted that may affect how the incoming audio is interpreted. These include acoustic conditions, such as background noise and echo, and network conditions, such as packet loss. In this paper, we focus on packet-loss telemetry as a representative environmental signal. Individual observations may be transient and do not by themselves indicate whether an impairment is ongoing or has recovered. \Sys therefore aligns these observations with the received-audio timeline and preserves evidence of both active impairment and subsequent recovery. This temporal state helps the context orchestrator distinguish environmental effects from user behavior when interpreting the current audio.

\subsection{Context Orchestration}
\label{sec:context-orchestration}

\begin{figure}[t]
    \centering
    \includegraphics[width=\columnwidth]{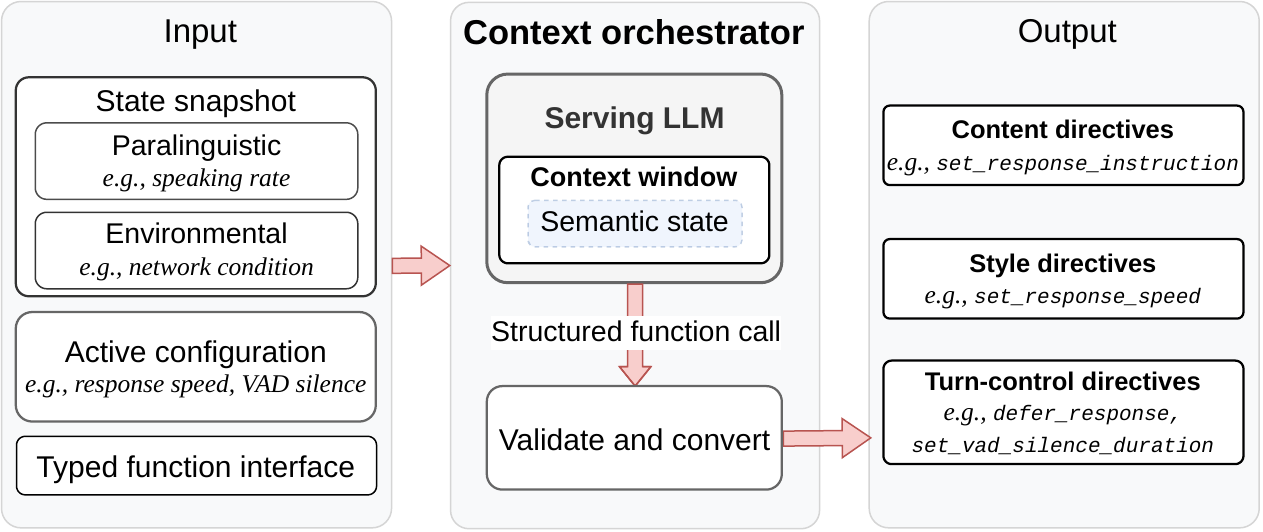}
    \caption{Context orchestrator maps the current states and runtime configuration to validated runtime directives.}
    \label{fig:context-orchestration}
\end{figure}
Given the voice context constructed
above, the state-to-control path determines how its semantic, paralinguistic, and environmental states jointly affect runtime control. 
Applying separate rules to each state can produce conflicting actions because each rule considers only part of the current voice context. For example, a user may explicitly ask the system to speak faster while speaking at a slower pace. The semantic state favors following the explicit request, while the paralinguistic state favors matching the observed speaking rate. Similarly, semantic evidence may suggest that a turn is complete, while recent network impairment suggests that an apparent pause may not reflect the user's behavior.

To resolve these conflicts, \sys introduces a context orchestrator that uses the serving LLM for in-band, state-aware reasoning, as shown in Figure~\ref{fig:context-orchestration}. The same LLM used for response generation performs orchestration over the current voice context rather than delegating the decision to a separate model. Because the semantic state is already present in the model context window, the serving LLM can reuse the current input and projected history without serializing and synchronizing them with another out-of-band controller. It augments this semantic context with the current paralinguistic and environmental states so that the LLM can interpret the three states jointly.

At each turn, the context orchestrator provides the serving LLM with three additional inputs. A \emph{state snapshot} contains the current paralinguistic and environmental states. The \emph{runtime configuration} records the current values of the available actuators, such as response speed and the active turn-detection configuration. A \emph{typed function interface} defines the controls available to the LLM and constrains the type and valid range of each argument. The state snapshot and runtime configuration provide the information needed to choose runtime controls, while the typed interface prevents the LLM from issuing arbitrary commands to the LLM-serving layer.

When control is needed, the serving LLM returns one or more structured function calls. The context orchestrator validates the calls and converts valid calls into runtime directives for the LLM-serving layer. The directives cover three types of control: content directives modify the instructions used for the next response; style directives adjust supported delivery attributes; and turn-control directives
determine turn-taking behavior (\eg whether response generation should be deferred and the VAD silence duration used for subsequent user speech). At each turn, the context orchestrator may produce no directive, one directive, or multiple directives. For example, an environmental state indicating packet loss may lead to a turn-control directive that defers the response and adapts turn detection. At the same time, if packet loss causes information to be missing from the received content, the context orchestrator may also issue a content directive asking the user to repeat or confirm the missing information.

The context orchestration overhead is conditional. If the current states do not require adjustment, the serving LLM does not return a function call, and response generation proceeds with the existing configuration. When orchestration invokes control functions, it introduces additional model usage cost and latency, but it may avoid more expensive downstream actions, such as generating an unnecessary response after a premature turn boundary. We evaluate this conditional overhead in \S\ref{sec:evaluation}.

\subsection{Structured Voice Memory}
\label{sec:structured-history}

\begin{figure}[t]
    \centering
    {\includegraphics[width=\columnwidth]{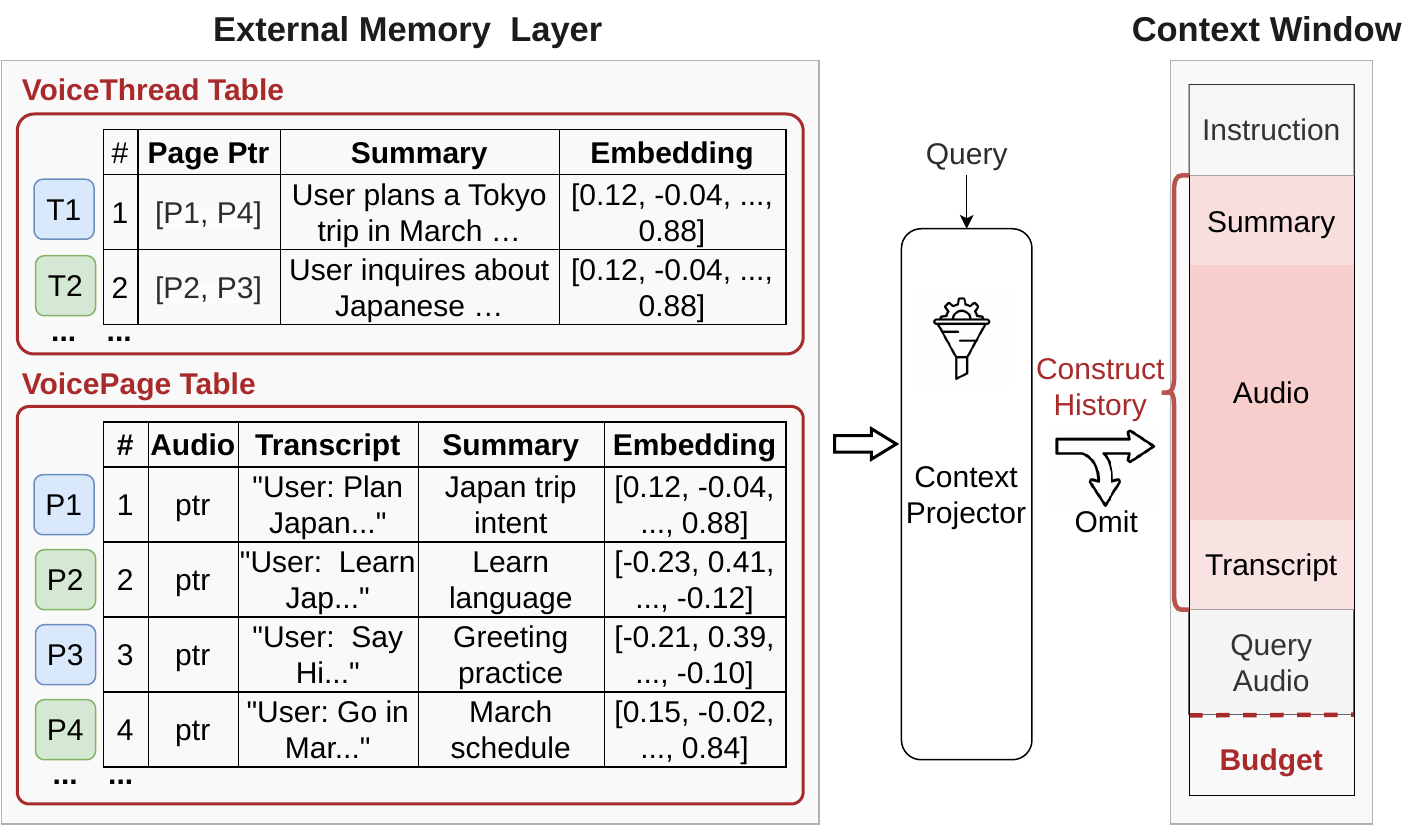}}
    \caption{Memory layout and context-window construction. The context projector uses \texttt{vthreads} and \texttt{vpages} to construct projected history under a context budget, selecting among audio, transcript, and summary representations.}
    \label{fig:systemmem}
\end{figure}

Constructing a bounded context first requires recovering conversational structure from a continuous, unstructured stream of voice interactions. 
At the micro level, a voice session consists of discrete interactions (\ie user--LLM exchanges). At the macro level, these aggregate into coherent conversational flows around topics or goals, with future interactions selectively referencing subsets of these flows.
Interactions within the same flow need not be temporally adjacent: a user may suspend one task, discuss another topic, and later return to the original task through an implicit spoken reference. A flat sequence cannot represent these two levels of structure, forcing the system either to retrieve isolated fragments or to include unrelated interactions. To represent both levels explicitly, \sys introduces two memory abstractions: a \texttt{vpage} represents a bounded interaction, while a \texttt{vthread} organizes related \texttt{vpages} into a coherent conversational flow, as shown in Figure~\ref{fig:systemmem}.

\pb{VoicePage.}
A \texttt{vpage} is the persistent unit for one completed voice interaction,
typically comprising a user input and the corresponding model output. It
records the exchange together with its available
representations, including references to audio, transcripts, summaries,
and the metadata needed for retrieval. The retained audio preserves
speech-delivery information, while the transcript and summary provide
progressively more compact semantic representations.

\pb{VoiceThread.}
Retrieving \texttt{vpages} independently can miss dependencies among interactions within the same conversational flow. A \texttt{vthread} addresses this problem by maintaining an ordered set of references to \texttt{vpages} associated with the same topic or goal. For example, a ``Trip Planning'' thread can link a hotel discussion with travel dates established earlier, allowing a later request such as ``Can you book that hotel?'' to recover both pieces of information. Thread membership is many-to-many: a \texttt{vthread} may reference multiple \texttt{vpages}, and a \texttt{vpage} may belong to multiple \texttt{vthreads} when one interaction constrains several topics or tasks. \texttt{Vthreads} store references rather than duplicating the audio and derived representations contained in their pages.

\pb{Thread Operations.}
Two operations organize a newly committed \texttt{vpage}. \textit{Append} associates
the page with matching
existing threads, whereas \textit{Spawn} creates a new thread after page-level
retrieval when no thread matches.

After committing a \texttt{vpage}, \sys retrieves relevant \texttt{vthreads} and appends the page to every matching thread. 
Supporting multiple matches is necessary because a single utterance may affect several ongoing tasks. For example, confirming a hotel for Tuesday while scheduling a client dinner for the same evening constrains both travel planning and meeting scheduling.

If no \texttt{vthread} matches, \sys performs page-level retrieval over the complete \texttt{vpage} index. Relevant historical pages are combined with the new page to spawn a new \texttt{vthread}. If no prior page matches, \sys creates a singleton \texttt{vthread} containing only the new page. This fallback handles topic drift and recovers details omitted by thread-level representations. For example, a food preference mentioned during a travel discussion can later seed a ``Dining Recommendation'' thread. Thread-level retrieval therefore preserves conversational continuity, while page-level retrieval recovers finer-grained details.

\pb{Maintenance.} %
Updating a \texttt{vthread}'s summary and retrieval representation on the latency-critical voice path would delay subsequent interactions. To avoid this delay, \sys makes each new \texttt{vpage} immediately available for page-level retrieval while asynchronously refreshing the affected \texttt{vthread}'s representations; once refreshed, they serve subsequent thread-level retrieval. Separating page availability from thread refresh removes thread-level aggregation from the latency-critical path while keeping the latest interaction accessible. Details of page and thread construction and retrieval are provided in \S\ref{sec:memory-context-implementation}.

\subsection{Multi-Fidelity Context Projection}
\label{sec:context-projection}

Structured voice memory makes accumulated interaction history retrievable, but each model request remains subject to a bounded context budget. Constructing the context window therefore requires selecting only the history relevant to the current interaction. Furthermore, not all relevant history needs to be retained as audio. A transcript or summary may suffice. The system must therefore determine both which history to include and the fidelity at which to represent it.

A common approach is to compact history according to recency, preserving recent interactions at high fidelity while compressing or discarding older ones. This approach assumes recent history is more relevant. Long-running voice conversations, however, frequently branch across topics and later return to earlier goals. An older interaction may contain the exact constraint needed by the current request, while several recent interactions may be unrelated. Uniformly compressing older history can therefore remove useful detail, whereas retaining all recent history can waste the context budget on irrelevant content.

To address this problem, \sys introduces a context projector that jointly selects relevant history and its representation fidelity, as shown in Figure~\ref{fig:systemmem}. The projector first retrieves candidate \texttt{vthreads} and \texttt{vpages} for the current request, then selects a representation for each candidate subject to the available context budget.

\pb{Candidate retrieval.}
Given the current request \(q\), the context projector first retrieves relevant \texttt{vthreads}. A matched thread provides a compact thread-level representation together with references to the \texttt{vpages} that form its conversational flow. Thread-level retrieval preserves related constraints that may not individually resemble the current request. For example, retrieving a ``Trip Planning'' thread for a hotel-booking request also makes the earlier page containing the travel dates available for projection.

If no \texttt{vthread} matches the request, the projector falls back to page-level retrieval. This fallback recovers relevant details that may not be retained in thread-level representations. \texttt{Vpages} referenced by multiple candidate threads are identified by their page identifiers and considered only once. The result is a set of candidate \texttt{vthreads} and \texttt{vpages} from which the projector constructs the historical context.

\pb{Representation fidelity.}
Each candidate can be represented as \emph{audio}, \emph{transcript}, or \emph{summary}, trading representation fidelity for model-input cost. Audio retains original acoustic and semantic information, a transcript retains lexical content without speech delivery, and a summary retains only compact information about topic and task. For a \texttt{vpage}, the projector selects one of its stored representations. A \texttt{vthread} can enter the context as a compact thread summary or be expanded into its referenced \texttt{vpages}. When expanded, a \texttt{vthread} is replaced by its referenced \texttt{vpages}, which are then scheduled independently. 
Let \(B\) denote the total context budget. The projector reserves an output safety buffer \(B_{\mathrm{safe}}\) and uses
$
B_{\mathrm{ctx}} = B - B_{\mathrm{safe}}
$
for historical context. The current request is retained as audio.

\pb{Policy.}
A natural formulation would select one representation \(m_i\) for each candidate \(u_i\) to maximize the relevance-weighted information retained in the context while satisfying
$
\sum_i c_{i,m_i} \leq B_{\mathrm{ctx}},
$
where \(c_{i,m_i}\) is the model-input cost of the selected representation. Solving this formulation exactly is impractical for real-time voice interaction: global optimization introduces additional runtime latency, and precisely quantifying the absolute information gain of a specific modality is intractable.

To resolve this, \sys uses a relevance-ordered layered-degradation heuristic. The projector initially assigns each candidate its highest available fidelity. If the projected context exceeds \(B_{\mathrm{ctx}}\), it processes candidates in increasing order of relevance and progressively degrades their representations along
$
\text{audio}
\rightarrow
\text{transcript}
\rightarrow
\text{summary}
\rightarrow
\text{omit}.
$
The projector skips representations that are unavailable and stops once the context satisfies the budget. This heuristic requires only an ordering over representation fidelity rather than an absolute estimate of information gain, while allowing more relevant interactions to retain higher-fidelity information when the budget permits.
Finally, the context projector serializes the selected representations in conversational order. These representations form the projected history used for context orchestration and response generation.

The policy has $O(N \log N)$ time complexity, where $N$ is the number of candidate units. 
Our empirical evaluations show that the scheduling overhead is around 1~ms, making it suitable for real-time applications. 

\section{\Sys Implementation}
\label{sec:runtime}

The \sys prototype comprises 1,246 lines of code (LoC), excluding model prompt templates, and includes a browser-based WebRTC media path with a Python control and context-management backend. The receiver runs in a Chromium-based browser and uses the browser's native WebRTC/Opus stack and NetEq implementation for media decoding and packet-loss concealment. 
The receiver collects WebRTC telemetry and forwards NetEq-decoded audio to the backend as 24-kHz PCM16 audio, preserving packet-loss concealment artifacts.
A Python asynchronous bridge relays the PCM stream to the OpenAI Realtime API over WebSocket, processes VAD and response events, executes orchestration decisions, and issues runtime VAD updates. Persistent conversation metadata and vector indexes are maintained in PostgreSQL with \texttt{pgvector} and HNSW-based retrieval.
\Sys can be ported to other model APIs that support streaming I/O and incremental context updates, such as Gemini Live and GLM-Realtime.
Open-source packages and demos for \sys are available at \url{https://llmovoice.com}.

\subsection{State-to-Control Path}
\label{sec:state-control-implementation}

\pb{Paralinguistic state extraction.}
For speaking rate, the runtime divides the transcript word count of each completed utterance by its audio duration. It discards utterances shorter than 0.5 seconds and, for English speech, observations outside 100--300 words per minute. The estimator retains the ten most recent valid observations. Let \(r_1,\ldots,r_n\) denote these observations in chronological order. The current speaking-rate estimate is computed using a linearly weighted average:
$\hat{r}_t =
\frac{\sum_{i=1}^{n} i r_i}{\sum_{i=1}^{n} i}.$
The linear weights smooth turn-level variation while allowing the estimate to follow sustained changes in the user's speaking rate.

\pb{Environmental state extraction.}
The receiver samples WebRTC reception statistics every 50\,ms and aggregates them over a rolling window of length \(W=\max\{1000\,\mathrm{ms},D_{\mathrm{VAD}}\}\), where \(D_{\mathrm{VAD}}\) is the current silence threshold used for turn detection. The resulting state records the total duration of concealed audio, the duration of concealment rendered as silence, and the elapsed time since the most recent concealment. The first two measurements characterize the severity and form of recent packet loss, while the last indicates whether the impairment remains recent. 

\pb{Orchestration.} %
The prototype runs joint state reasoning through an instruction that guides the serving LLM to generate control decisions based on the voice context. When the environmental state indicates that the received input may be incomplete or unreliable, the instruction prioritizes obtaining a complete input before response generation. The LLM may therefore defer the response or ask the user to repeat missing information. Otherwise, an explicit request in the semantic state takes precedence over a delivery preference inferred from the paralinguistic state. For example, a request to ``speak faster'' overrides a slower speaking rate observed from the user. Subject to these principles, the LLM selects and coordinates controls for response content, speaking rate, response deferral, and future turn detection. This instruction guides context-dependent decisions without prescribing a fixed mapping from individual states to controls.

\pb{Directive execution.}
The runtime exposes these controls through four typed functions. \texttt{set\_\allowbreak response\_\allowbreak instruction} adds a content instruction to the next response, and \texttt{set\_\allowbreak response\_\allowbreak speed} updates the provider-supported speaking-rate setting. \texttt{defer\_\allowbreak response} prevents immediate response generation while the current turn boundary remains provisional, whereas \texttt{set\_\allowbreak vad\_\allowbreak silence\_\allowbreak duration} updates the VAD silence threshold for subsequent turn boundaries. Before applying a function call, the runtime control adapter verifies that the requested control is supported and that its arguments satisfy the declared types and ranges. Invalid calls are rejected without changing the runtime configuration. 

\pb{Style directives.}
When orchestration determines that the response delivery should be adapted (\eg to align with the user's observed speaking rate or follow an explicit request), it invokes \texttt{set\_\allowbreak response\_\allowbreak speed}. 
Let
$s_{\min}$ and $s_{\max}$ be the provider's minimum and maximum speed
settings, which profiling maps to response rates $r_{\min}$ and $r_{\max}$.
We map user rate $r$ to
\[
s(r)=\operatorname{clip}\!\left(
s_{\min}+\frac{r-r_{\min}}{r_{\max}-r_{\min}}
(s_{\max}-s_{\min}),\,s_{\min},\,s_{\max}\right).
\]
For the OpenAI Realtime API used in our prototype, speed settings
$s_{\min}=0.8$ and $s_{\max}=1.5$ yield
$r_{\min}=142.4$ and $r_{\max}=267.0$ words per minute, respectively.

\pb{Turn-control directives.}
When the environmental state indicates that an apparent turn boundary may result from network impairment rather than the end of the user's turn, orchestration may invoke \texttt{defer\_\allowbreak response} to defer the current response. A deferred response is reconsidered after a 1.5-second soft deadline. If the network has recovered, the runtime generates the response. Otherwise, it continues waiting until a 5-second hard deadline forces response generation.
If a \texttt{speech\_started} event arrives while a response is pending, the runtime treats the new speech as a continuation of the current interaction. It cancels the pending response, waits for the corresponding \texttt{speech\_stopped} event, and generates one response over the combined input. This continuation confirms that the preceding VAD boundary was premature. The runtime therefore uses the inter-segment gap to adjust the VAD silence duration for subsequent turns. 
Let $g$ be the gap between the preceding
\texttt{speech\_stopped} and following \texttt{speech\_started} events.
With safety margin $\delta$, update
\[
D_{\mathrm{new}}=
\min\!\left(D_{\max},
q\left\lceil\frac{D_{\mathrm{VAD}}+g+\delta}{q}\right\rceil\right).
\]
Here $q$ is the VAD granularity and $D_{\max}$ is the maximum threshold. Our prototype uses $q=100\,\mathrm{ms}$, $\delta=200\,\mathrm{ms}$, and
$D_{\max}=5000\,\mathrm{ms}$, and restores the
default after five consecutive seconds of healthy reception.

When the current semantic and environmental states indicate that future turn boundaries require adapting the silence threshold, orchestration invokes \texttt{set\_\allowbreak vad\_\allowbreak silence\_\allowbreak duration} proactively. In contrast, the automatic VAD update above in \texttt{defer\_\allowbreak response} requires no LLM function call and occurs only after subsequent speech confirms that a boundary was premature.

\subsection{Memory-to-Context Path}
\label{sec:memory-context-implementation}

The prototype stores \texttt{vpage} and \texttt{vthread} metadata in PostgreSQL and maintains their many-to-many relationships in a separate membership table. Audio payloads reside in the local file system, with their paths recorded in the corresponding \texttt{vpage} records. For content processing and retrieval, \sys uses \textit{gpt-4o-mini-transcribe} for transcription, \textit{gpt-4o-mini} for summarization, and a locally executed \textit{BAAI/bge-small-en-v1.5} model to generate 384-dimensional retrieval embeddings. The embeddings are stored in a \texttt{pgvector} hierarchical navigable small world (HNSW) index~\cite{malkov2018efficient}; relevance is computed using cosine similarity. Retrieval embeddings for \texttt{vpages} and \texttt{vthreads} are generated from their respective summaries, while the transcript of the current request serves as the retrieval query. Representation availability is determined from the stored metadata.

To estimate model-input costs, the projector applies the provider-specific pricing model to estimated audio and text token usage. Audio tokens are estimated from the total duration of the user and model-response audio, whereas text tokens for transcripts and summaries are estimated from their word counts. When selected \texttt{vthreads} are expanded into page-level representations, the runtime deduplicates shared \texttt{vpages} by identifier and sums the costs of the selected page representations. When a compact thread-level summary is selected instead, its cost is estimated directly from its word count.

\section{Evaluation} \label{sec:evaluation}

We evaluate \sys by answering five research questions:
\begin{enumerate}[leftmargin=*, nosep]
    \item Does the system improve (a) speaking-rate alignment and (b) turn handling under network loss? (\S\ref{sec:voice-controls})
    \item Can the system bound the cost of processing long conversation histories? (\S\ref{sec:cost_scalability})
    \item Can the system accurately retrieve historical information and preserve answer quality? (\S\ref{sec:context_efficacy})
    \item What are the sources of the system's latency overhead, and how much does each contribute? (\S\ref{sec:system_overhead})
    \item How does the system perform in real-world use? (\S\ref{sec:application_case_studies})
\end{enumerate}

Our key findings are: (1) \textbf{Voice-specific control}: For RQ1, 
speaking-rate control reduces mean
absolute error (MAE) by 52.4\%.
Under packet loss, turn-taking control reduces the false-interruption rate (FIR) by 98.0\% and network-induced cost by 79.2\%. (2) \textbf{Model usage cost}: For RQ2, \sys keeps per-turn cost bounded and is up to 24.9$\times$ cheaper than OpenAI (and still 1.8$\times$ cheaper than the idealized 100\%--cached OpenAI case) in long-context conversations. (3) \textbf{Context quality}: For RQ3, \sys achieves 86.9\%/75.3\% retrieval hit rates and 7.7/6.3 judge scores on RT2/MSP-PODCAST, outperforming the evaluated bounded-context alternatives in answer quality. (4) \textbf{Overhead}: For RQ4, context organization adds 78.4\,ms (3.8\%) on RT2 and 50.9\,ms (2.7\%) on MSP-PODCAST; turns that produce a control function call incur a separate mean decision latency of 790.2\,ms. (5) \textbf{Robustness}: For RQ5, our case studies show that explicit voice context and persistent memory improve interaction quality by supporting topic resumption, network-aware turn handling, and recovery from disconnections.

\subsection{Experimental Setup}
\label{sec:eval-setup}

\textbf{Workloads.}
We use three conversational-speech workloads.
(1) \textit{NIST Rich Transcription 2002 (RT2)}: RT2 contains 60 real-world telephone conversations of
approximately five minutes each~\cite{2002RichTranscription2004}.
(2) \textit{MSP-PODCAST}: MSP-PODCAST contains longer recordings with diverse
speakers and acoustic conditions; we select 70 sessions with at least two
speakers and 30 turns, averaging 58 minutes~\cite{Busso_2025}.
(3) \textit{ICASSP PLC}: The validation set of the ICASSP 2024 Audio Deep Packet Loss
Concealment (PLC) Challenge contains 800 clean 48-kHz recordings of natural human
speech and packet-loss traces sampled from loss patterns observed in real
Microsoft Teams user calls at 20-ms granularity.
We use RT2 for the speaking-rate experiment in RQ1(a), the ICASSP PLC validation
set for the packet-loss experiment in RQ1(b), and RT2 and MSP-PODCAST for the
long-context experiments in RQ2--RQ3.

\pb{Benchmark.}
For RQ2--RQ3, we construct the long-context benchmark by first ingesting each RT2 and MSP-PODCAST session through the same runtime pipeline, which constructs \texttt{vpages} and \texttt{vthreads} and stores transcripts, embeddings,
and metadata. From each processed session, we sample a partial history and select a target fact whose evidence appears in one \texttt{vpage} (\textit{single-page}) or spans multiple \texttt{vpages} (\textit{multi-page}). For each fact, we generate a question and reference answer, synthesize the question using \texttt{tts-1}, and use the resulting audio as the query. The resulting benchmark contains 120 RT2 and 138 MSP-PODCAST queries.

\pb{Experimental platform.}
The experiments use the prototype implementation described in
\S\ref{sec:runtime}, including its transcription, summarization,
embedding, storage, and retrieval stack. We use \texttt{gpt-realtime-mini}
for response generation and GPT-4o as the answer-quality judge. 

\pb{Compared methods.}
For RQ1(a), we compare against \textit{\sys w/o context orchestration}, which uses the default response speed. 
For RQ1(b), we compare against: 
(1) \textit{Fixed-500}, which disables context orchestration and fixes
\texttt{silence\_duration\_ms} at 500\,ms;
(2) \textit{Static-Max}, which uses a conservative but non-adaptive threshold
of 3200\,ms.

For RQ2--RQ3, we compare: 
(1) \textit{OpenAI (Full History)}, which directly uses the default OpenAI Realtime API with full-history prompts, serving as an unconstrained upper bound on accuracy. We also compare prompt-cached variants. However, cache hit rates cannot be expected to reach 100\% in practice, since in-memory cached prefixes typically
persist for only a short time (\eg 5--10 minutes of inactivity)~\cite{OpenAIPromptCaching}.
(2) \textit{Page-RAG}, which retrieves at the page level using
embedding similarity under the same cost budget as ours,
representing a standard retrieval-based baseline without context structuring. 
(3) \textit{Compaction}, which follows a commonly adopted implementation: after each assistant response, if the accumulated context exceeds
the same cost budget as ours, earlier history is summarized
while retaining only the most recent two turns verbatim~\cite{ContextSummarizationRealtime}. This represents a practical compression-based baseline used
in real-world LLM systems.
(4) \textit{Hybrid RAG}, which combines the latest two turns with relevance- and
recency-based page retrieval. It unions the relevance and recency of the top-10
candidates and ranks them by
$0.7\times\text{relevance}+0.3\times\text{recency}$.
(5) \textit{Hierarchical Summary}, which combines the latest two turns with maintained
summaries of older history, grouping older interactions into 10-turn chunks. 

In our figures, \textit{RAG}, \textit{Comp.}, \textit{H-RAG}, \textit{H-Sum}, and \textit{LOV} denote Page-RAG, Compaction, Hybrid RAG, Hierarchical Summary, and \sys, respectively.

\pb{Controlled experimental protocol.}
For RQ1(a), both conditions receive the same query audio, selected context,
model, prompt, and voice. The experiment changes only whether context
orchestration maps the observed WPM state to the next response speed; it
therefore measures speaking-rate alignment rather than subjective naturalness.
For RQ1(b), each ICASSP packet-loss mask is applied to a fresh WebRTC session in Chromium. NetEq decodes the received stream, and all decoded PCM (including
packet-loss-concealed audio) is forwarded to Realtime. Fixed-500, Static-Max,
and \sys receive the same source recording and loss mask, and each
trace-condition pair uses a fresh Realtime session.

\pb{Parameter settings.}
Thread- and page-level retrieval use a cosine-similarity threshold of 0.35.
All bounded-context methods use the same context budget within each
comparison. Projection estimates audio at 9.98 tokens per second and text at 1.33 tokens per word based on profiling of the model's input accounting. 
For VAD settings, \sys initializes VAD silence threshold $D_{\mathrm{VAD}}$ to 500\,ms.

\pb{Metrics.}
For RQ1(a), we use \textit{WPM Alignment Error (MAE)}, the mean absolute difference between user and response WPM
($\frac{1}{N}\sum_{i=1}^{N}|\mathrm{WPM}^{(i)}_{\mathrm{response}}-\mathrm{WPM}^{(i)}_{\mathrm{user}}|$).
For RQ1(b), we report:
(1) \textit{False-Interruption Rate (FIR)}, the fraction of traces in which a premature endpoint produces audible response audio
($\frac{\#\text{traces with false responses}}{\#\text{evaluated traces}}$);
(2) \textit{Endpoint Response Latency}, measured from the final endpoint event to the first response audio
($t^{(i)}_{\mathrm{first\ audio}}-t^{(i)}_{\mathrm{final\ endpoint}}$); and
(3) \textit{Network-Induced Cost}, the average per-trace cost of false responses and orchestration
($C_{\mathrm{network}}=\frac{C_{\mathrm{false\ responses}}+C_{\mathrm{orchestration}}}{N}$).

For RQ3, we use:
(1) \textit{Retrieval Hit Rate}, the fraction of ground-truth \texttt{vpages} loaded into the final context; and 
(2) \textit{LLM-as-a-Judge Score}, for which GPT-4o rates semantic correctness from 1 to 10 using the same prompt across methods. Semantically equivalent answers are accepted regardless of wording differences. 
For RQ4, we use \textit{time
to first audio (TTFA)}, measured from the end of the user query to the first response-audio token. Because the new stronger-baseline runs use a different real-time proxy, their TTFA is reported only descriptively and is not used for cross-experiment latency comparisons.

\subsection{Runtime Control with Voice Context}
\label{sec:voice-controls}

\textbf{Paralinguistic-aware speaking-rate control.}
Without context orchestration, the model responds at its default speaking rate, resulting in a WPM MAE of 38.26. With context orchestration enabled, \sys maps the observed speaking rate encoded in the paralinguistic state to a response-speed directive, reducing the MAE to 18.23. This corresponds to an absolute reduction of 20.03 WPM and a relative reduction of 52.4\%. The result shows that paralinguistic-aware control improves speaking-rate alignment, although it does not, by itself, establish greater subjective naturalness.

\begin{table}[t]
\centering
\caption{Turn-control results over ICASSP PLC traces.}
\label{tab:environment-control}
\small
\begin{tabular}{lrrr}
\toprule
\textbf{Policy} &
\textbf{FIR} &
\textbf{Latency (ms)} &
\textbf{Network cost (\$/trace)} \\
\midrule
Fixed-500
  & 46.0\%
  & 1286.1
  & 0.003344 \\
Static-Max
  & 0.0\%
  & 4021.4
  & 0 \\
\sys
  & 0.9\%
  & 1830.7
  & 0.000695 \\
\bottomrule
\end{tabular}
\end{table}

\pb{Environment-aware turn control.}
\Cref{tab:environment-control} reports turn-control effectiveness over the traces from the validation set of
the ICASSP 2024 Audio Deep Packet Loss Concealment
Challenge. 
Fixed-500 produced audible premature responses on 46.0\% of traces,
whereas \sys reduced FIR to 0.9\%, a 98.0\% relative reduction.
Static-Max also eliminated false interruptions, but its conservative
3.2-s silence threshold increased median endpoint response latency to
4.02\,s, compared with 1.29\,s for Fixed-500 and 1.83\,s
for \sys. Finally, Fixed-500 incurred \$0.003344 per trace in costs
from premature responses. In comparison, \sys reduced the total
network-induced cost to \$0.000695 per trace, a 79.2\% reduction.
This total comprises \$0.000069 in residual false-response costs and
\$0.000626 in function-call costs for context orchestration.

\begin{figure}[t]
    \centering
    \subfloat[RT2]{\includegraphics[width=0.5\columnwidth]{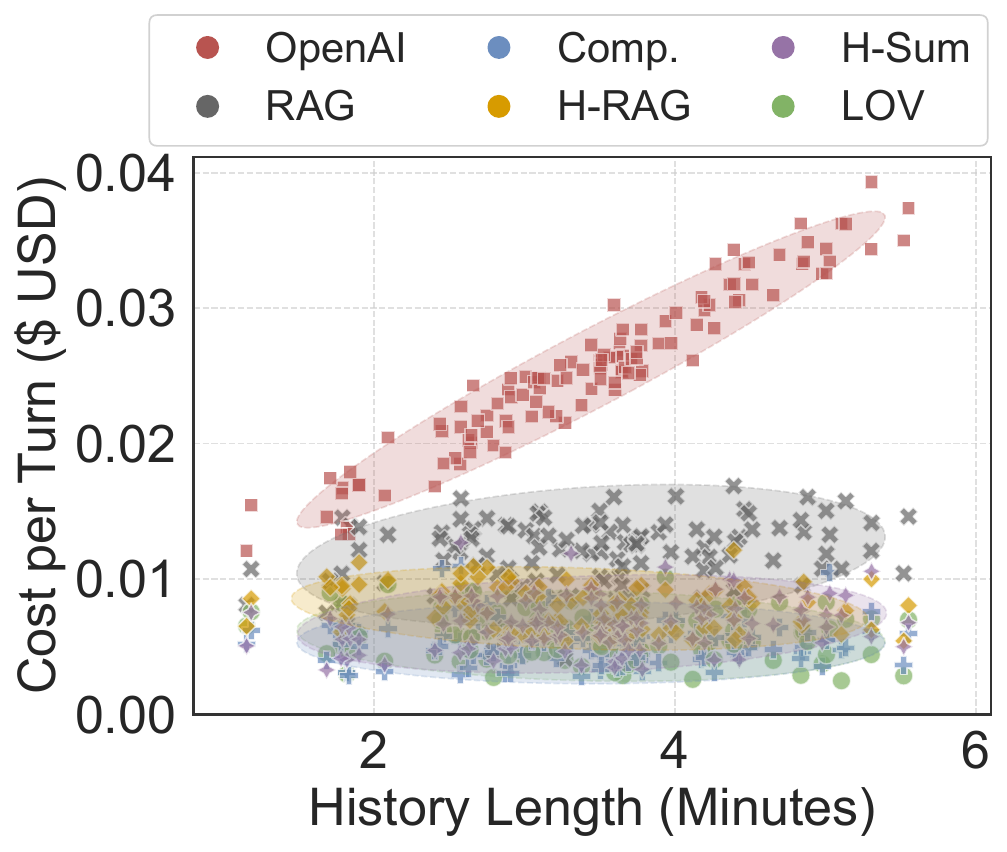}}\hfill
    \subfloat[MSP-PODCAST]{\includegraphics[width=0.5\columnwidth]{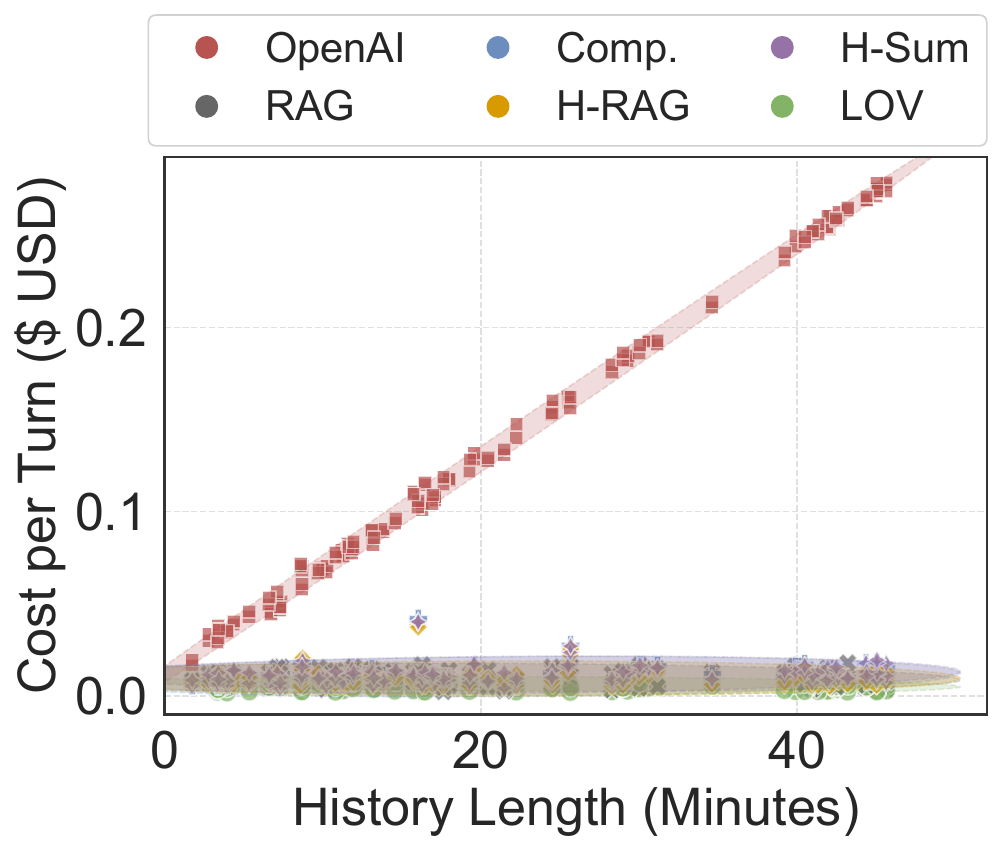}}
    \caption{Cost per turn versus conversation history length.}
    \label{fig:cost_vs_length}
\end{figure}
\begin{figure}[t]
    \centering
    \subfloat[RT2]{\includegraphics[width=0.5\columnwidth]{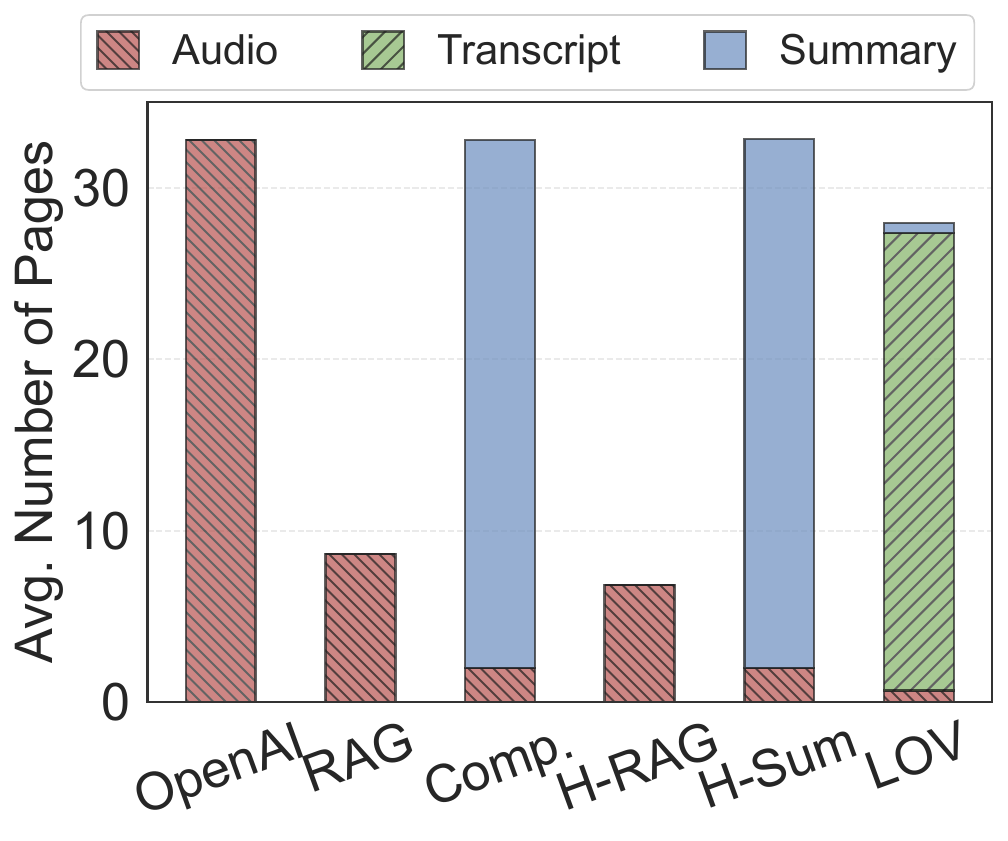}}\hfill
    \subfloat[MSP-PODCAST]{\includegraphics[width=0.5\columnwidth]{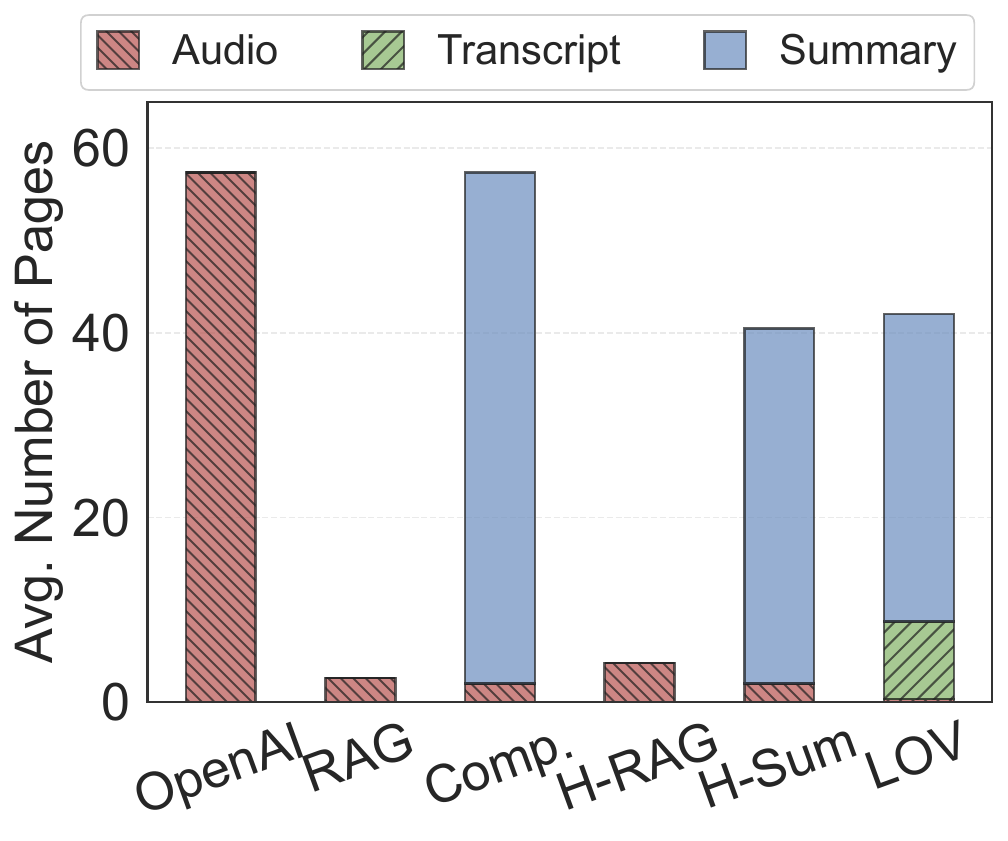}}
    \caption{Page composition in the context window.}
    \label{fig:context_usage}
\end{figure}

\begin{table}[t]
    \centering
    \caption{Average per-turn cost comparison.}
    \vspace{-0.25em}
    \label{tab:avg_cost_comparison}
    \footnotesize
    \begin{tabular}{lrr}
        \toprule
        \textbf{Method} & \textbf{RT2} & \textbf{MSP-PODCAST} \\
        \midrule
        OpenAI & \$0.025468 & \$0.140198 \\
        \quad 25\% cached & \$0.020346 & \$0.107614 \\
        \quad 50\% cached & \$0.015224 & \$0.075029 \\
        \quad 75\% cached & \$0.010103 & \$0.042445 \\
        \quad 90\% cached & \$0.007029 & \$0.022894 \\
        \quad 100\% cached & \$0.004981 & \$0.009860 \\
        Page-RAG & \$0.011753 & \$0.010112 \\
        Compaction & \$0.005297 & \$0.010543 \\
        Hybrid RAG & \$0.007821 & \$0.009487 \\
        Hierarchical Summary & \$0.006686 & \$0.011446 \\
        \sys & \textbf{\$0.005799} & \textbf{\$0.005634} \\
        \bottomrule
    \end{tabular}
\end{table}

\subsection{Cost Scalability}\label{sec:cost_scalability}
As shown in Table~\ref{tab:avg_cost_comparison}, \sys remains cost-efficient on both workloads. On RT2, \sys is 4.4$\times$ cheaper than OpenAI. On MSP-PODCAST, \sys is 24.9$\times$ cheaper than OpenAI. Notably, \sys remains competitive even under strong OpenAI caching (\eg 90\%--100\%). On long conversations (MSP-PODCAST), \sys is still 1.8$\times$ cheaper than the idealized 100\%-cached OpenAI case. 

Figure~\ref{fig:cost_vs_length} illustrates the scalability of \sys. As preloaded conversation history length grows (measured by cumulative user-audio duration, since the OpenAI API preloads only user audio while agent history is loaded as transcripts), OpenAI exhibits a steep upward cost trend due to expanding context windows. In contrast, \sys remains flat and strictly bounded across history lengths. 
This advantage comes from the context projector described in \S\ref{sec:context-projection}. As shown in Figure~\ref{fig:context_usage}, compared to OpenAI, which keeps expanding heavy raw-audio context, our projector explicitly downgrades less relevant pages into lightweight transcript and summary representations as history grows. Representation maintenance is also cheap: transcription and summarization account for only 4.6\% (\$0.000269/query) of \sys's full cost on RT2 and 5.1\% (\$0.000285/query) on MSP-PODCAST. Local embedding, search, and storage incur no additional external API cost, and their latency is reported in \S\ref{sec:system_overhead}. 

\subsection{Context Efficacy}\label{sec:context_efficacy}
Under the same budget, the bounded-context baselines achieve costs comparable to \sys. The primary question is whether their selected context is equally effective.
\begin{figure}[t]
    \centering
    \subfloat[RT2]{\includegraphics[width=0.5\columnwidth]{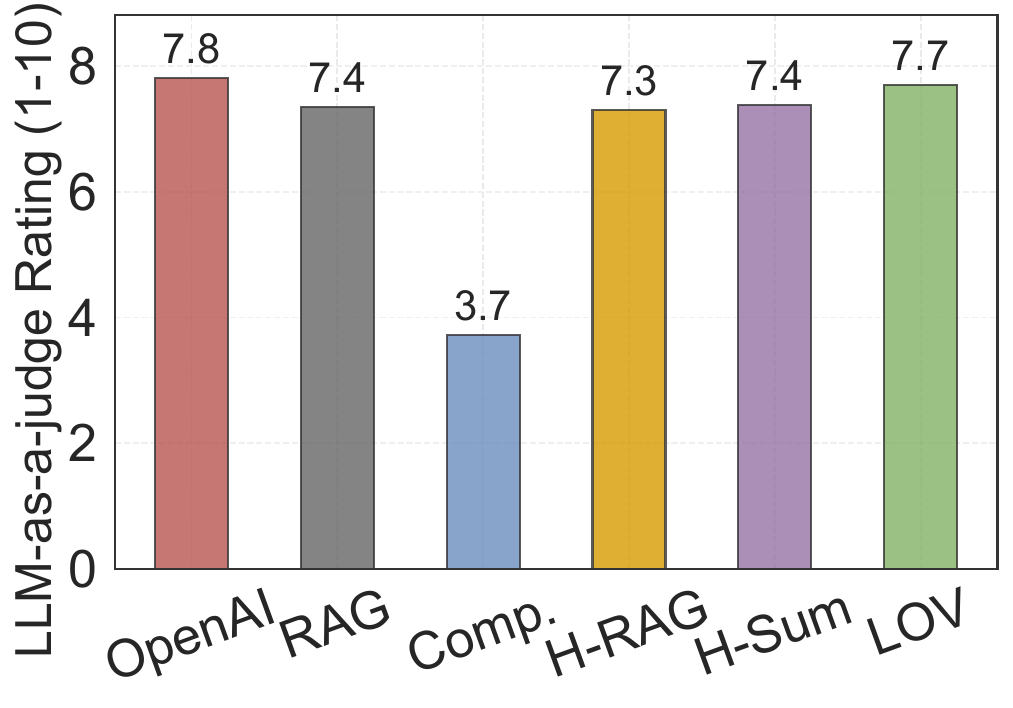}}\hfill
    \subfloat[MSP-PODCAST]{\includegraphics[width=0.5\columnwidth]{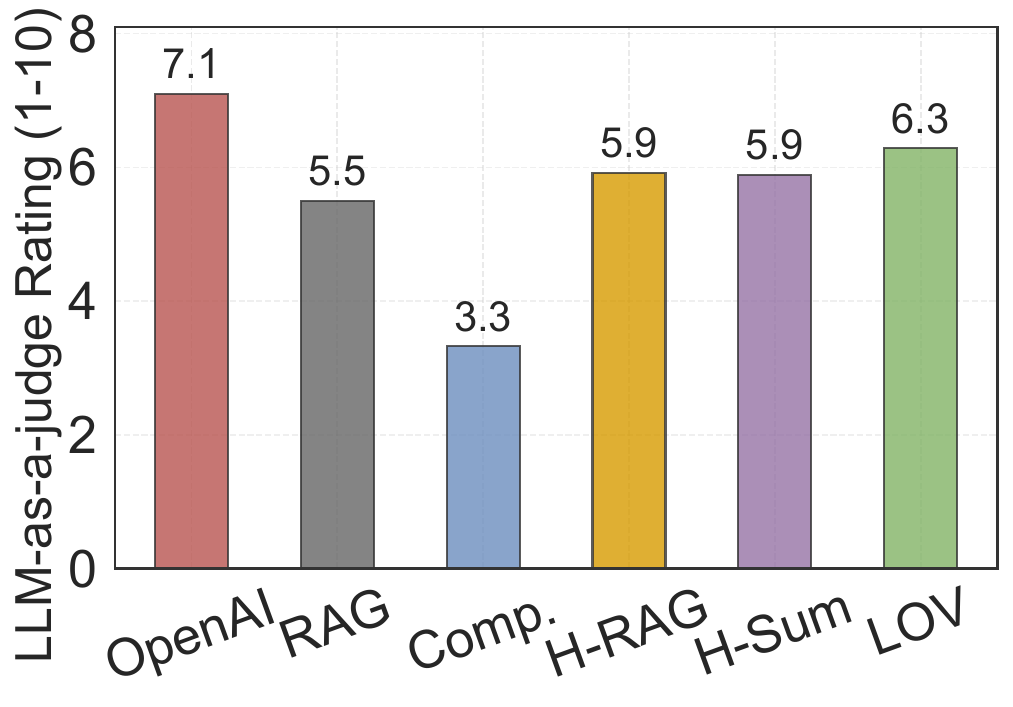}}
    \caption{LLM-as-a-Judge answer-quality comparison.}
    \label{fig:llm_judge}
\end{figure}
We first examine LLM-as-a-Judge ratings as shown in Figure~\ref{fig:llm_judge}. \Sys preserves the best answer quality compared to RAG and Compaction and remains close to OpenAI: \sys retains 98.7\% and 88.7\% of OpenAI-level answer quality on RT2 and MSP-PODCAST, respectively. 
Both Compaction and Hierarchical Summary achieve lower answer quality than \sys across both workloads. This is due to their static policies, which retain recent turns verbatim while compressing older history into summaries that can omit fine-grained details. In contrast, \sys performs dynamic, relevance-aware compaction at the thread and page levels. As shown in Figure~\ref{fig:context_usage}, it retains more high-fidelity context when affordable and gradually shifts to compact representations as sessions grow longer. For example, compared with RT2, \sys represents more of the older context as summaries rather than transcripts on the longer MSP-PODCAST sessions. However, this compression can also cause accuracy loss. We observe a 14.6\% transcript-to-summary failure rate, counting a failure only when the answer is incorrect and the required fact is inferable from the full transcript but not from the loaded summary.

\begin{figure}[t]
    \centering
    \subfloat[RT2]{\includegraphics[width=0.5\columnwidth]{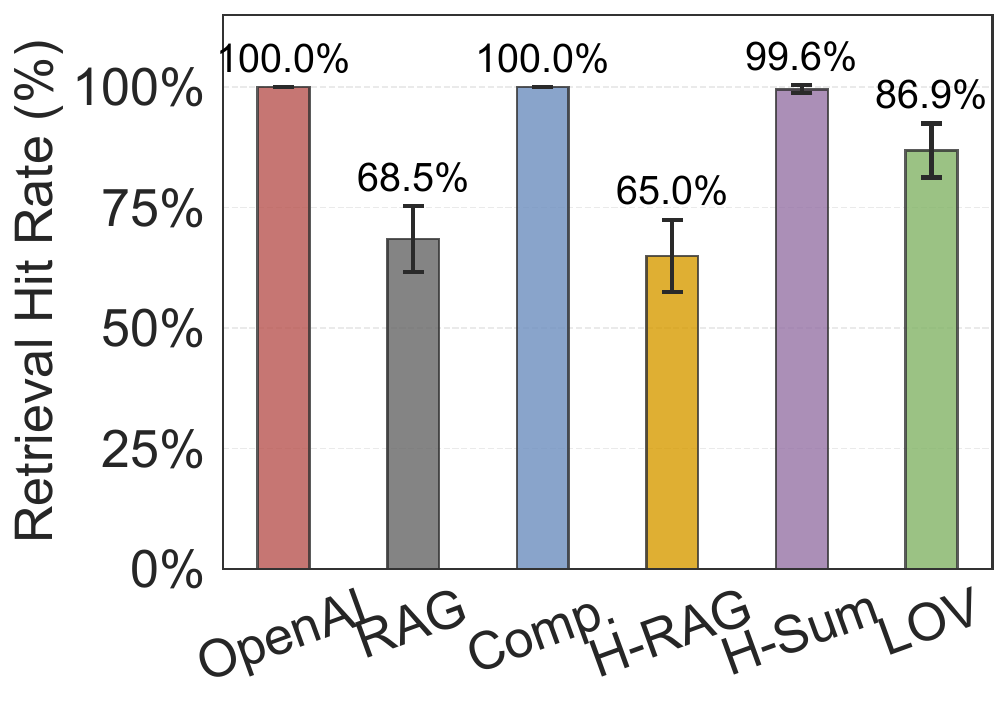}}\hfill
    \subfloat[MSP-PODCAST]{\includegraphics[width=0.5\columnwidth]{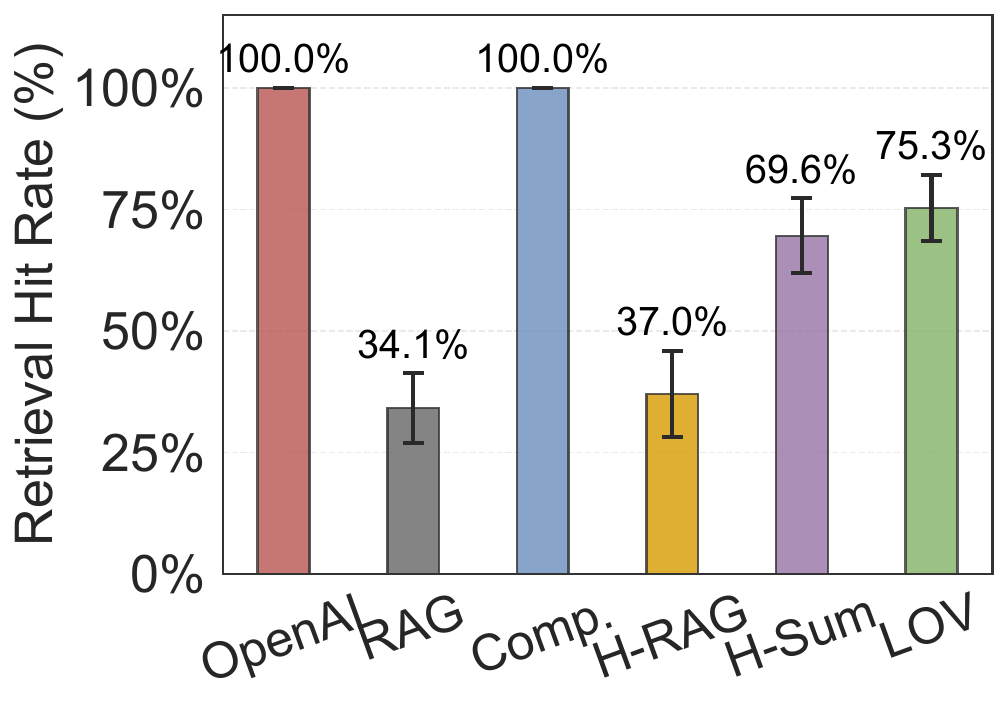}}
    \caption{Retrieval hit-rate comparison on both workloads.}
    \label{fig:hit_rate}
\end{figure}
Compared with \sys, Page-RAG retains 3.8\% and 11.3\% less OpenAI-level answer quality on RT2 and MSP-PODCAST, respectively. The corresponding quality gaps for Hybrid RAG are 4.5\% and 4.8\%. To better understand this gap, we next examine retrieval hit rate in Figure~\ref{fig:hit_rate}. Both Page-RAG and Hybrid RAG have substantially lower retrieval accuracy, especially on the longer MSP-PODCAST conversations. This limitation is particularly pronounced for multi-page queries. For example, the multi-page request hit rate of Page-RAG drops by 26.4\% on RT2 (81.7\% to 55.3\%) and 27.4\% on MSP-PODCAST (47.8\% to 20.4\%) relative to single-page queries. This reflects a shared locality failure: page-level retrieval breaks cross-page conversational locality, so evidence distributed across turns is not jointly retrieved. In contrast, \sys preserves this locality by organizing \texttt{vpages} into coherent \texttt{vthreads}, enabling thread-level retrieval of connected evidence.

\subsection{System Overhead}\label{sec:system_overhead}
\begin{figure}[t]
    \centering
    \subfloat[RT2]{\includegraphics[width=0.47\columnwidth]{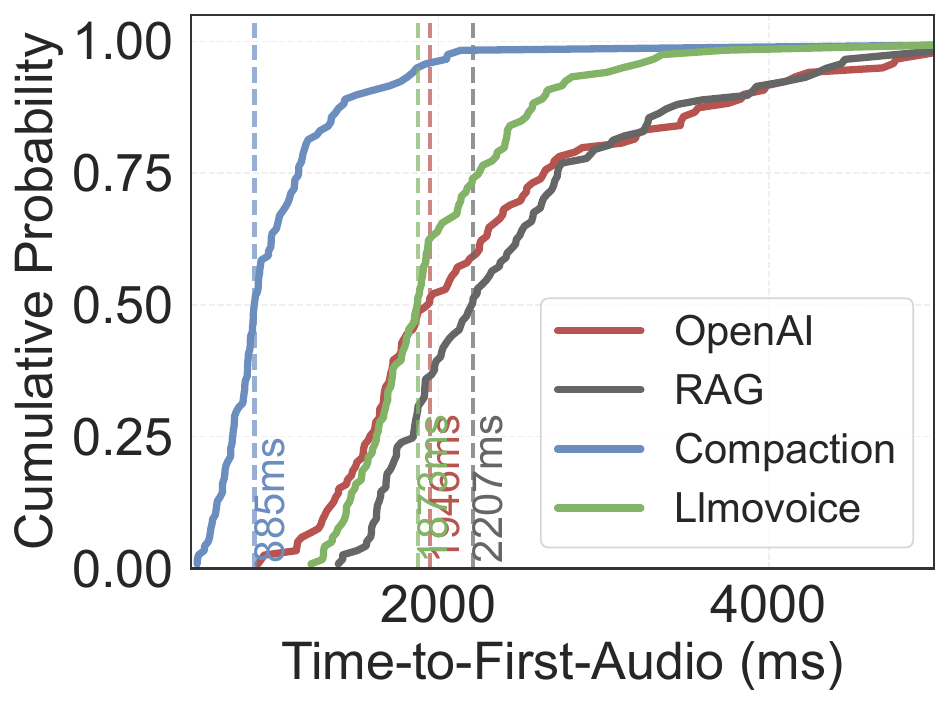}}
    \hfill
    \subfloat[MSP-PODCAST]{\includegraphics[width=0.5\columnwidth]{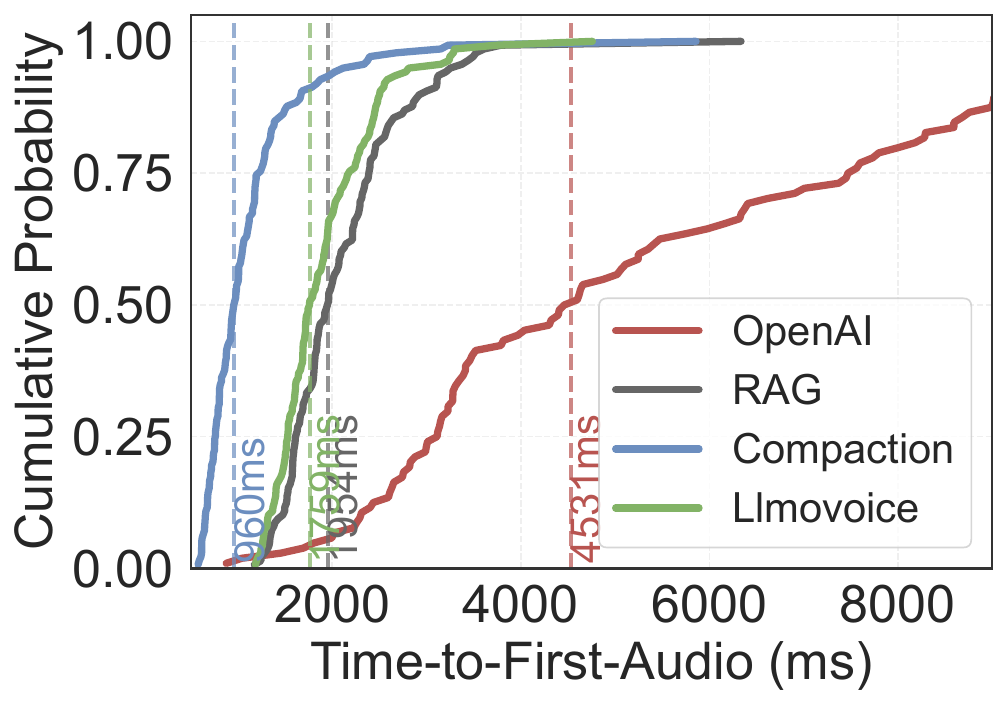}}
    \caption{Cumulative distribution of TTFA by workload.}
    \label{fig:ttfa_cdf}
\end{figure}

We next evaluate the latency overhead introduced by the middleware layer. 
We first measure the memory-to-context path using end-to-end TTFA. These long-context runs exercise context retrieval and projection without producing a function call. Table~\ref{tab:ttfa_general} compares \sys with three representative baselines across both workloads.
\begin{table}[t]
    \centering
     \caption{End-to-end TTFA by workload and method.}
     \vspace{-0.25em}
    \label{tab:ttfa_general}
    \small
    \begin{tabular}{l l r r r}
    \toprule
    \textbf{Workload} & \textbf{Method} & \textbf{TTFA (ms)} & \textbf{95\% CI} & \textbf{n} \\
    \midrule
    RT2 & OpenAI & 2451 & $\pm$339 & 120 \\
    RT2 & Page-RAG & 2644 & $\pm$264 & 120 \\
    RT2 & Compaction & 1066 & $\pm$137 & 120 \\
    RT2 & \sys & 2070 & $\pm$156 & 120 \\
    \midrule
    MSP & OpenAI & 7510 & $\pm$864 & 138 \\
    MSP & Page-RAG & 2102 & $\pm$113 & 138 \\
    MSP & Compaction & 1117 & $\pm$105 & 138 \\
    MSP & \sys & 1909 & $\pm$92 & 138 \\
    \bottomrule
    \end{tabular}
\end{table}
Figure~\ref{fig:ttfa_cdf} shows the corresponding latency distributions.
The CDFs further show a clear rightward shift for OpenAI on
both workloads, especially on MSP-PODCAST. In contrast,
the \sys curves rise earlier and more steeply, indicating
that a larger fraction of queries reach first audio with lower
latency and less dispersion. OpenAI also exhibits a heavier
high-latency tail, whereas \sys shows shorter tails.

To identify latency sources, we provide a component-level breakdown in Table~\ref{tab:latency_breakdown}.
\begin{table}[t]
    \centering
    \caption{Latency breakdown for \sys{} (Gen: generation, Emb: embedding,
    Vec: vector search, PJ: context projection).}
    \label{tab:latency_breakdown}
    \small
    \begin{tabular}{l r r r r}
    \toprule
    \multirow{2}{*}{\textbf{Component}}
        & \multicolumn{2}{c}{\textbf{RT2}}
        & \multicolumn{2}{c}{\textbf{MSP-PODCAST}} \\
    \cmidrule(lr){2-3}\cmidrule(lr){4-5}

        & \multicolumn{1}{c}{\textbf{Lat. (ms)}}
        & \multicolumn{1}{c}{\multirow{1}{*}{\textbf{\%}}}
        & \multicolumn{1}{c}{\textbf{Lat. (ms)}}
        & \multicolumn{1}{c}{\multirow{1}{*}{\textbf{\%}}} \\

    \midrule
    STT   & 525.4  & 25.4\%  & 490.3  & 25.7\% \\
    Gen   & 1465.4 & 70.8\%  & 1367.5 & 71.7\% \\
    Emb   & 60.6   & 2.9\%   & 28.9   & 1.5\%  \\
    Vec   & 17.4   & 0.8\%   & 20.9   & 1.1\%  \\
    PJ    & 0.3    & 0.0\%   & 1.1    & 0.1\%  \\
    Input & 0.1    & 0.0\%   & 0.0    & 0.0\%  \\
    \midrule
    \textbf{Total}
        & \textbf{2069.1}
        & \textbf{100.0\%}
        & \textbf{1908.7}
        & \textbf{100.0\%} \\
    \bottomrule
    \end{tabular}
\end{table}
Table~\ref{tab:latency_breakdown} shows that the memory-to-context overhead of \sys remains lightweight in both workloads: 78.4\,ms (3.8\%) on RT2 and 50.9\,ms (2.7\%) on MSP-PODCAST. Most of this internal time comes from embedding and vector retrieval, while context projection and input construction are negligible. In contrast, external STT and generation account for the dominant share of end-to-end TTFA, indicating that the memory-to-context path contributes only a minor fraction of total latency. Notably, our current implementation performs embeddings over STT transcripts, while audio-based embeddings are also feasible. This can bypass STT waiting time and further reduce TTFA in practice.

The state-to-control path incurs a separate, conditional overhead when the serving LLM within the context orchestrator generates a function call. We measure decision latency from the submission of the state snapshot to the receipt of the structured function call. Among turns that generated such a call, the mean decision latency was 790.2\,ms (median: 698.8\,ms; 95th percentile: 1,286.6\,ms).

\subsection{Evaluating \Sys in the Real World}\label{sec:application_case_studies}

To examine dynamic interactions between users and voice applications in real-world settings, %
we deployed \sys to handle open-ended travel-planning tasks. In this experiment, each volunteer interacted with \sys, \textsf{OpenAI Native (Realtime API)}, and \textsf{OpenAI with Context Compaction} in randomized order. In each session, the volunteer stated two constraints (\eg location preference), switched to two subtasks, and then returned with an additional constraint and requested a final recommendation. We report three representative cases from these unscripted interactions.

\pb{Case \#1: Maintaining detailed information across topic drifts.} 
A volunteer is planning a trip and asks for tourist attraction recommendations. In Phase 1, they set two constraints: \textit{"I want a niche, natural landscape with very few people."} After several minutes of unrelated chitchat (Phase 2), the user returns to the topic in Phase 3 and adds: \textit{"For the attractions you just recommended, I want a place with convenient transportation."}

\pb{Observation:} 
In the \textsf{OpenAI with Context Compaction} baseline, the system suffers from "detail amnesia." Because of the sub-task chitchat, the background summarizer compressed the early history and completely forgot the "few people" constraint. It confidently recommended \textit{Dali Ancient City}, stating: \textit{"Dali has great transportation with high-speed rail and an airport."} This recommendation violates the user's requirement, as Dali is highly crowded. \textsf{OpenAI Native} retained all constraints by processing the full history, but cost \$0.54/min.

In contrast, when the user returned to the topic, \sys's thread-level retrieval precisely fetched the prior \texttt{vthread} containing the tourist-attraction constraint. 
It successfully recommended \textit{Shuhe Ancient Town}: \textit{"You can consider Shuhe; it is quiet, uncrowded, and relatively easy to reach."} Because \sys projects only the relevant threads (in transcript) rather than the full audio history, it provided the correct answer at a cost of only \$0.13/min.

\pb{Case \#2: Preventing false triggers under natural network fluctuations.} During the conversation, the volunteer experiences a temporary network fluctuation, causing unpredictable packet loss and audio delivery delays.

\pb{Observation:}
We observed a catastrophic failure mode in the baselines. Under network fluctuation, the baseline's static VAD (with the default \texttt{silence\_duration\_ms}=200\, ms) misinterpreted network-induced audio gaps as the user pausing or finishing their sentence. Consequently, the baseline continuously interrupted the user with fragmented, nonsensical responses: \textit{"Sure, no problem... Please go on... Understood, I will... Wait a moment... Okay, I'll give you..."} 
Because the baseline blindly feeds these false-triggered responses back into the context window, it rapidly accumulated an extra 7,206 wasted tokens in a very short time. This not only drove up the per-turn cost for the remainder of the session but also caused the agent to hallucinate based on fragmented audio until the user manually corrected the flow.
In \sys, the network condition is explicitly represented in the environmental states. The context orchestrator reasons over the states and issues a VAD control directive, so delayed packets are awaited instead of prematurely triggering generation. This explicit state-driven control reduces false interruptions and avoids cascading token waste.

\pb{Case \#3: State recovery upon disconnection.}
 During severe network degradation, the assistant's response audio became choppy and discontinuous, and TTFA rose to an intolerable level. After noticing the prolonged delay, a volunteer proactively hung up and reconnected.

\pb{Observation:}
In our experiment, the OpenAI Realtime baseline did not preserve session state across reconnection and therefore treated the reconnected interaction as a new session. When the volunteer asked, \textit{"Where were we?"}, the baseline replied, \textit{"Hello! Nice to meet you! How can I help?"} The user had to explain again, \textit{"I just asked you to recommend food."} The baseline then asked, \textit{"Got it. You want me to recommend food again, right? Tell me what you like."} This forced the user to repeat previously provided information. In contrast, \sys stores prior turns independently of the active connection. When the volunteer reconnected and asked, \textit{"Where were we?"}, \sys retrieved the active \texttt{thread} and resumed the conversation from its pre-disconnection state: \textit{"Hello! Nice to see you again! Shall we continue with the food recommendation, or do you want to talk about something else?"}

\section{Related Work}\label{sec:related_work}

\noindent\textbf{LLM-over-voice systems.}
Early voice conversational stacks mostly follow cascaded designs: input speech is transcribed to text, processed by an LLM, and then converted back to speech~\cite{huang2024audiogpt,shen2023hugginggpt}. To reduce response latency and retain speech information discarded during transcription, recent efforts shift toward end-to-end, real-time speech-native models~\cite{defossez2024moshi,du2023lauragpt,chen2025slam,nguyen2023generative,fang2024llama,zhang2025omniflatten,RealtimeAPIOpenAI,GeminiLiveAPI,xu2025qwen3}. However, this line of work primarily optimizes model-centric properties, such as speech quality, response latency, and duplex interaction, under controlled settings. Production LLM-over-voice systems must additionally maintain cross-turn states about how utterances are delivered and conditions under which audio arrives, and coordinate them with semantic context to control the live voice pipeline. Existing model architectures do not expose such a persistent serving state. \Sys fills this gap by providing explicit voice states and context orchestration.

\noindent\textbf{Persistent conversational memory.}
Long-term-memory systems extend an LLM beyond its immediate context window by storing and recovering prior information. LongMem~\cite{wang2023augmenting} augments a frozen language model with a retrievable memory network. MemoryBank~\cite{zhong2024memorybank} continuously updates conversational memories and user profiles. MemGPT~\cite{packer2023memgpt} moves information between in-context and external memory tiers. Mem0~\cite{chhikara2025mem0} extracts, consolidates, and retrieves salient facts from multi-session conversations. Collectively, these systems show that supporting long-term interaction requires a persistent memory substrate separate from the model's bounded working context. \Sys adopts this principle for voice interaction through \texttt{vpage}, which stores each completed interaction together with its available representations and retrieval embeddings. This persistent record preserves both semantic content and speech-delivery evidence for downstream organization, retrieval, and orchestration.

\noindent\textbf{RAG.}
RAG grounds generation by retrieving relevant passages from an external corpus~\cite{lewis2020retrieval, guu2020retrieval}. Flat RAG~\cite{karpukhin2020dense} treats passages independently, whereas hierarchical RAG retrieves information across multiple levels of abstraction. RAPTOR~\cite{sarthi2024raptor} recursively clusters and summarizes passages into a retrieval tree, while GraphRAG~\cite{edge2024graphrag} and HippoRAG~\cite{gutierrez2024hipporag} use graphs to connect evidence across a corpus. These methods are designed around document passages, entities, or facts. In contrast, \texttt{vthread} captures the temporal evolution of voice conversations by organizing ordered \texttt{vpages} into coherent threads based on their voice characteristics, enabling hierarchical retrieval.

\noindent\textbf{Context reduction.}
Context reduction techniques, such as summarization, context compaction, and windowing, trade representation detail for a smaller model input. LLMLingua ~\cite{jiang2023llmlingua} and LongLLMLingua~\cite{jiang2024longllmlingua} remove less informative prompt tokens under a compression budget, while RECOMP~\cite{xu2024recomp} learns extractive and abstractive compressors for retrieved passages and can omit retrieval when it adds no useful information. These techniques operate primarily on text and optimize which semantic content survives compression. Projection extends this idea to voice history: under a fixed context budget, it represents each exchange using audio, a transcript, or a summary, or omits it entirely.

\noindent\textbf{Caching.}
Serving systems use prompt, prefix, and KV caching to reduce repeated context-processing cost by reusing previously computed model states. Prompt Cache precomputes attention states for reusable prompt modules~\cite{gim2024promptcache}. vLLM's PagedAttention manages KV-cache memory efficiently and enables sharing across requests~\cite{kwon2023pagedattention}. CacheBlend~\cite{yao2025cacheblend} combines cached states when retrieved knowledge is assembled into a new prompt. These techniques optimize the processing of constructed context and complement \sys's context organization.

Compared with these lines of work, \sys combines two capabilities that they do not jointly provide. Its state-to-control path goes beyond speech-native models by explicitly representing semantic, paralinguistic, and environmental states and jointly mapping them to runtime controls. Its memory-to-context path extends text-centric memory, RAG, and context-reduction techniques to voice interactions by organizing persistent, temporally ordered records and selecting among audio, transcript, summary, and omission under a context budget. Caching optimizes an already constructed context, whereas \sys determines what voice context to construct. Together, these new capabilities make long-running voice interaction more scalable while adapting system behavior to maintain interaction quality. 

\section{Conclusion}
\label{sec:conclusion}

This paper presents \sys, a voice-context management middleware that explicitly models semantic, paralinguistic, and environmental states in voice interaction. Through a voice-context orchestration loop, \sys jointly reasons over these states to derive runtime controls and constructs bounded context from structured interaction units represented at multiple fidelities, rather than from monolithic full history. Across benchmarks and end-to-end cases, \sys improves speaking-rate alignment and robustness to network fluctuations while reducing long-session costs and preserving response quality.

\begin{acks}
We thank the anonymous reviewers and our shepherd for their insightful comments. The work conducted at Shanghai Jiao Tong University was supported by the National Key R\&D Program of China (Grant No.~2024YFC3017100) and the National Natural Science Foundation of China (Grant No.~62302292). We also acknowledge technical support from AgenticSys Group (\url{https://agenticsys.com}). Corresponding author: Yifei Zhu (\nolinkurl{yifei.zhu@sjtu.edu.cn}).
\end{acks}

\bibliographystyle{ACM-Reference-Format}
\bibliography{llmovoice}

@misc{VoiceFirstGenerativeAI,
  author      = {{GSMA}},
  title = {Voice-{{First Generative AI}} for {{Impact}}: {{Insights}} from {{Viamo}}'s {{Ask Viamo Anything}} Pilot in {{Zambia}}},
  journal = {Mobile for Development},
  urldate = {2026-08-29},
  langid = {british},
  year = {2025},
  url = {https://www.gsma.com/solutions-and-impact/connectivity-for-good/mobile-for-development/wp-content/uploads/2025/03/GSMA_Viamo_Voice-First-Generative-AI-for-Impact_2025.pdf}

}

@article{medhi2011designing,
author = {Medhi, Indrani and Patnaik, Somani and Brunskill, Emma and Gautama, S.N. Nagasena and Thies, William and Toyama, Kentaro},
title = {Designing mobile interfaces for novice and low-literacy users},
year = {2011},
volume = {18},
number = {1},
journal = {ACM Trans. Comput.-Hum. Interact.},
month = may,
articleno = {2},
numpages = {28},
}

@article{malkov2018efficient,
  title={Efficient and robust approximate nearest neighbor search using hierarchical navigable small world graphs},
  author={Malkov, Yu A and Yashunin, Dmitry A},
    journal = {IEEE Trans. Pattern Anal. Mach. Intell.},
    volume={42},
  number={4},
  pages={824--836},
  year={2018},
  publisher={IEEE}
}

@inproceedings{chen2026speechmedassist,
  title={SpeechMedAssist: Efficiently and Effectively Adapting Speech Language Models for Medical Consultation},
  author={Chen, Sirry and Wang, Jieyi and Chen, Wei and Wei, Zhongyu},
  booktitle = {Proc. ACL},
  pages={30914--30935},
  year={2026}
}

@article{chang2026vrcoaching,
  title={Designing and evaluating LLM-driven coaching agents for VR public speaking practice},
  author={Chang, Enyao and Che, Xiaoping and Zhang, Jianing},
  journal = {Front. Psychol.},
  volume={17},
  pages={1886780},
  year={2026},
  publisher={Frontiers Media SA}
}

@article{high2026artificial,
  title={Artificial intelligence for agricultural extension: Supporting transformative learning among smallholder farmers},
  author={High, Chris and Singh, Namita and Nemes, Guszt{\'a}v},
  journal={J. Dev. Policy Pract.},
  volume={11},
  number={1},
  pages={61--80},
  year={2026},
  publisher={Sage Publications Sage India: New Delhi, India}
}

@article{monk2026invehicle,
  title={Visual and Cognitive Demands of a Large Language Model-Powered In-vehicle Conversational Agent},
  author={Monk, Chris and Ayala, Allegra and Yu, Christine SP and Fitch, Gregory M and Gruber, Dara},
  journal={arXiv preprint arXiv:2601.15034},
  year={2026}
}

@article{diener2025icassp,
  title={The ICASSP 2024 audio deep packet loss concealment grand challenge},
  author={Diener, Lorenz and Branets, Solomiya and Saabas, Ando and Cutler, Ross},
  journal={IEEE Open J. Signal Process.},
  volume={6},
  pages={231--237},
  year={2025},
  publisher={IEEE}
}

@inproceedings{jiang2023active,
  title={Active retrieval augmented generation},
  author={Jiang, Zhengbao and Xu, Frank F and Gao, Luyu and Sun, Zhiqing and Liu, Qian and Dwivedi-Yu, Jane and Yang, Yiming and Callan, Jamie and Neubig, Graham},
  booktitle={Proc. EMNLP},
  pages={7969--7992},
  year={2023}
}

@article{borrie2019syncing,
  title={Syncing up for a good conversation: A clinically meaningful methodology for capturing conversational entrainment in the speech domain},
  author={Borrie, Stephanie A and Barrett, Tyson S and Willi, Megan M and Berisha, Visar},
  journal={J. Speech Lang. Hear. Res.},
  volume={62},
  number={2},
  pages={283--296},
  year={2019},
  publisher={American Speech-Language-Hearing Association}
}

@article{brennan1996lexical,
  title={Lexical entrainment in spontaneous dialog},
  author={Brennan, Susan E and others},
  journal={Proc. ISSD},
  volume={96},
  pages={41--44},
  year={1996}
}

@article{2002RichTranscription2004,
  title={2002 rich transcription broadcast news and conversational telephone speech},
  author={Garofolo, J and Fiscus, Jonathan and Le, Audrey},
  journal={Linguistic Data Consortium},
  year={2004}
}

@article{garrod1987saying,
  title={Saying what you mean in dialogue: A study in conceptual and semantic co-ordination},
  author={Garrod, Simon and Anderson, Anthony},
  journal={Cognition},
  volume={27},
  number={2},
  pages={181--218},
  year={1987},
  publisher={Elsevier}
}

@misc{ContextSummarizationRealtime,
	title = {Context {Summarization} with {Realtime} {API}},
	url = {https://developers.openai.com/cookbook/examples/context_summarization_with_realtime_api},
  author = {{OpenAI}},
  year = {2026},
  note = {(accessed Apr. 2, 2026)}
}

@article{Busso_2025,
  author  = {Busso, Carlos and Lotfian, Reza and Sridhar, Kusha
             and Salman, Ali N. and Lin, Wei-Cheng and Goncalves, Lucas
             and Parthasarathy, Srinivas and Naini, Abinay Reddy
             and Leem, Seong-Gyun and Martinez-Lucas, Luz
             and Chou, Huang-Cheng and Mote, Pravin},
  title   = {The {MSP-Podcast} Corpus},
  journal = {IEEE Trans. Affect. Comput.},
  pages   = {1--19},
  year    = {2026},
}

@inproceedings{fang2024llama,
  title={Llama-omni: Seamless speech interaction with large language models},
  author={Fang, Qingkai and Guo, Shoutao and Zhou, Yan and Ma, Zhengrui and Zhang, Shaolei and Feng, Yang},
  booktitle = {Proc. ICLR},
  volume={2025},
  pages={57607--57624},
  year={2025}
}

@inproceedings{cui2025recent,
  title={Recent advances in speech language models: A survey},
  author={Cui, Wenqian and Yu, Dianzhi and Jiao, Xiaoqi and Meng, Ziqiao and Zhang, Guangyan and Wang, Qichao and Guo, Steven Y and King, Irwin},
  booktitle={Proc. ACL},
  pages={13943--13970},
  year={2025}
}

@article{xu2025qwen3,
  title={Qwen3-omni technical report},
  author={Xu, Jin and Guo, Zhifang and Hu, Hangrui and Chu, Yunfei and Wang, Xiong and He, Jinzheng and Wang, Yuxuan and Shi, Xian and He, Ting and Zhu, Xinfa and others},
  journal={arXiv preprint arXiv:2509.17765},
  year={2025}
}

@inproceedings{zhang2025omniflatten,
  title={Omniflatten: An end-to-end gpt model for seamless voice conversation},
  author={Zhang, Qinglin and Cheng, Luyao and Deng, Chong and Chen, Qian and Wang, Wen and Zheng, Siqi and Liu, Jiaqing and Yu, Hai and Tan, Chao-Hong and Du, Zhihao and others},
  booktitle = {Proc. ACL},
  pages={14570--14580},
  year={2025}
}

@article{nguyen2023generative,
  title={Generative spoken dialogue language modeling},
  author={Nguyen, Tu Anh and Kharitonov, Eugene and Copet, Jade and Adi, Yossi and Hsu, Wei-Ning and Elkahky, Ali and Tomasello, Paden and Algayres, Robin and Sagot, Benoit and Mohamed, Abdelrahman and others},
  journal = {Trans. Assoc. Comput. Linguist.},
  volume={11},
  pages={250--266},
  year={2023},
  publisher={MIT Press One Broadway, 12th Floor, Cambridge, Massachusetts 02142, USA~…}
}

@inproceedings{guu2020retrieval,
  title={Retrieval augmented language model pre-training},
  author={Guu, Kelvin and Lee, Kenton and Tung, Zora and Pasupat, Panupong and Chang, Mingwei},
  booktitle={Proc. ICML},
  pages={3929--3938},
  year={2020},
  organization={PMLR}
}

@article{lewis2020retrieval,
  title={Retrieval-augmented generation for knowledge-intensive nlp tasks},
  author={Lewis, Patrick and Perez, Ethan and Piktus, Aleksandra and Petroni, Fabio and Karpukhin, Vladimir and Goyal, Naman and K{\"u}ttler, Heinrich and Lewis, Mike and Yih, Wen-tau and Rockt{\"a}schel, Tim and others},
  journal={Proc. NeurIPS},
  volume={33},
  pages={9459--9474},
  year={2020}
}

@article{defossez2024moshi,
  title={Moshi: a speech-text foundation model for real-time dialogue},
  author={D{\'e}fossez, Alexandre and Mazar{\'e}, Laurent and Orsini, Manu and Royer, Am{\'e}lie and P{\'e}rez, Patrick and J{\'e}gou, Herv{\'e} and Grave, Edouard and Zeghidour, Neil},
  journal={arXiv preprint arXiv:2410.00037},
  year={2024}
}

@article{du2023lauragpt,
  title={Lauragpt: Listen, attend, understand, and regenerate audio with gpt},
  author={Du, Zhihao and Wang, Jiaming and Chen, Qian and Chu, Yunfei and Gao, Zhifu and Li, Zerui and Hu, Kai and Zhou, Xiaohuan and Xu, Jin and Ma, Ziyang and others},
  journal={arXiv preprint arXiv:2310.04673},
  year={2023}
}

@inproceedings{chen2025slam,
  title={Slam-omni: Timbre-controllable voice interaction system with single-stage training},
  author={Chen, Wenxi and Ma, Ziyang and Yan, Ruiqi and Liang, Yuzhe and Li, Xiquan and Xu, Ruiyang and Niu, Zhikang and Zhu, Yanqiao and Yang, Yifan and Liu, Zhanxun and others},
  booktitle = {Findings of ACL},
  pages={2262--2282},
  year={2025}
}

@inproceedings{karpukhin2020dense,
  title={Dense passage retrieval for open-domain question answering},
  author={Karpukhin, Vladimir and Oguz, Barlas and Min, Sewon and Lewis, Patrick and Wu, Ledell and Edunov, Sergey and Chen, Danqi and Yih, Wen-tau},
  booktitle={Proc. EMNLP},
  pages={6769--6781},
  year={2020}
}

@article{packer2023memgpt,
  title={Memgpt: Towards llms as operating systems},
  author={Packer, Charles and Wooders, Sarah and Lin, Kevin and Fang, Vivian and Patil, Shishir G and Stoica, Ion and Gonzalez, Joseph E},
  journal={arXiv preprint arXiv:2310.08560},
  year={2023}
}

@inproceedings{huang2024audiogpt,
  title={Audiogpt: Understanding and generating speech, music, sound, and talking head},
  author={Huang, Rongjie and Li, Mingze and Yang, Dongchao and Shi, Jiatong and Chang, Xuankai and Ye, Zhenhui and Wu, Yuning and Hong, Zhiqing and Huang, Jiawei and Liu, Jinglin and others},
  booktitle = {Proc. AAAI},
  volume={38},
  number={21},
  pages={23802--23804},
  year={2024}
}

@article{shen2023hugginggpt,
  title={Hugginggpt: Solving ai tasks with chatgpt and its friends in hugging face},
  author={Shen, Yongliang and Song, Kaitao and Tan, Xu and Li, Dongsheng and Lu, Weiming and Zhuang, Yueting},
  journal={Proc. NeurIPS},
  volume={36},
  pages={38154--38180},
  year={2023}
}

@misc{StateMobileInternetb,
	title = {The State of Mobile Internet Connectivity 2024},
	url = {https://www.gsmaintelligence.com/research/the-state-of-mobile-internet-connectivity-2024},
  author = {GSMA},
  year = {2024},
  note = {(accessed Sep. 17, 2025)}
}

@misc{gubbala3InternetSmartphone2022,
  title        = {{Internet}, Smartphone and Social Media Use},
  author       = {Wike, Richard and Silver, Laura and Fetterolf, Janell and
                  Huang, Christine and Austin, Sarah and Clancy, Laura and
                  Gubbala, Sneha},
  year         = {2022},
  organization = {Pew Research Center},
  url          = {https://www.pewresearch.org/global/2022/12/06/internet-smartphone-and-social-media-use-in-advanced-economies-2022/},
  note         = {(accessed Aug. 19, 2025)}
}

@article{khan2024artificial,
  title={Artificial intelligence for low income countries},
  author={Khan, Muhammad Salar and Umer, Hamza and Faruqe, Farhana},
	journal = {Humanit. Soc. Sci. Commun.},
  volume={11},
  number={1},
  pages={1422},
  year={2024},
  publisher={Palgrave}
}

@misc{GeminiLiveAPI,
	title = {Gemini {Live} {API} Overview},
	url = {https://ai.google.dev/gemini-api/docs/live-api},
  author = {{Google}},
  year = {2025},
  note = {(accessed Mar. 30, 2026)}
}

@book{nass2005wired,
  title={Wired for speech: How voice activates and advances the human-computer relationship},
  author={Nass, Clifford Ivar and Brave, Scott},
  volume={9},
  year={2005},
  publisher={MIT press Cambridge, MA}
}

@misc{RealtimeAPIOpenAI,
	title = {Realtime and Audio},
	url = {https://developers.openai.com/api/docs/guides/realtime},
  author = {{OpenAI}},
  year = {2026},
  note = {(accessed Aug. 14, 2026)}
}

@misc{OpenAIPromptCaching,
  author = {{OpenAI}},
  title = {Prompt Caching},
  url = {https://developers.openai.com/api/docs/guides/prompt-caching},
  urldate = {2026-08-09},
  year = {2026},
  note = {(accessed Aug. 9, 2026)}
}

@article{chhikara2025mem0,
  title={Mem0: Building production-ready ai agents with scalable long-term memory},
  author={Chhikara, Prateek and Khant, Dev and Aryan, Saket and Singh, Taranjeet and Yadav, Deshraj},
  journal={arXiv preprint arXiv:2504.19413},
  year={2025}
}

@article{wang2023augmenting,
  title={Augmenting language models with long-term memory},
  author={Wang, Weizhi and Dong, Li and Cheng, Hao and Liu, Xiaodong and Yan, Xifeng and Gao, Jianfeng and Wei, Furu},
  journal={Proc. NeurIPS},
  volume={36},
  pages={74530--74543},
  year={2023}
}

@inproceedings{zhong2024memorybank,
  title={Memorybank: Enhancing large language models with long-term memory},
  author={Zhong, Wanjun and Guo, Lianghong and Gao, Qiqi and Ye, He and Wang, Yanlin},
  booktitle={Proc. AAAI},
  volume={38},
  number={17},
  pages={19724--19731},
  year={2024}
}

@inproceedings{sarthi2024raptor,
  title={Raptor: Recursive abstractive processing for tree-organized retrieval},
  author={Sarthi, Parth and Abdullah, Salman and Tuli, Aditi and Khanna, Shubh and Goldie, Anna and Manning, Christopher},
  booktitle = {Proc. ICLR},
  volume={2024},
  pages={32628--32649},
  year={2024}
}

@article{edge2024graphrag,
  title={From local to global: A graph rag approach to query-focused summarization},
  author={Edge, Darren and Trinh, Ha and Cheng, Newman and Bradley, Joshua and Chao, Alex and Mody, Apurva and Truitt, Steven and Metropolitansky, Dasha and Ness, Robert Osazuwa and Larson, Jonathan},
  journal={arXiv preprint arXiv:2404.16130},
  year={2024}
}

@article{gutierrez2024hipporag,
  title={Hipporag: Neurobiologically inspired long-term memory for large language models},
  author={Guti{\'e}rrez, Bernal J and Shu, Yiheng and Gu, Yu and Yasunaga, Michihiro and Su, Yu},
  journal={Proc. NeurIPS},
  volume={37},
  pages={59532--59569},
  year={2024}
}

@inproceedings{jiang2023llmlingua,
  title={Llmlingua: Compressing prompts for accelerated inference of large language models},
  author={Jiang, Huiqiang and Wu, Qianhui and Lin, Chin-Yew and Yang, Yuqing and Qiu, Lili},
  booktitle = {Proc. EMNLP},
  pages={13358--13376},
  year={2023}
}

@inproceedings{jiang2024longllmlingua,
  title={Longllmlingua: Accelerating and enhancing llms in long context scenarios via prompt compression},
  author={Jiang, Huiqiang and Wu, Qianhui and Luo, Xufang and Li, Dongsheng and Lin, Chin-Yew and Yang, Yuqing and Qiu, Lili},
  booktitle = {Proc. ACL},
  pages={1658--1677},
  year={2024}
}

@inproceedings{xu2024recomp,
  title={RECOMP: Improving retrieval-augmented LMs with context compression and selective augmentation},
  author={Xu, Fangyuan and Shi, Weijia and Choi, Eunsol},
  booktitle = {Proc. ICLR},
  volume={2024},
  pages={43478--43502},
  year={2024}
}

@inproceedings{gim2024promptcache,
  title={Prompt cache: Modular attention reuse for low-latency inference},
  author={Gim, In and Chen, Guojun and Lee, Seung-seob and Sarda, Nikhil and Khandelwal, Anurag and Zhong, Lin},
  booktitle = {Proc. MLSys},
  volume={6},
  pages={325--338},
  year={2024}
}

@inproceedings{kwon2023pagedattention,
  title={Efficient memory management for large language model serving with pagedattention},
  author={Kwon, Woosuk and Li, Zhuohan and Zhuang, Siyuan and Sheng, Ying and Zheng, Lianmin and Yu, Cody Hao and Gonzalez, Joseph and Zhang, Hao and Stoica, Ion},
  booktitle = {Proc. ACM SOSP},
  pages={611--626},
  year={2023}
}

@inproceedings{yao2025cacheblend,
  title={Cacheblend: Fast large language model serving for rag with cached knowledge fusion},
  author={Yao, Jiayi and Li, Hanchen and Liu, Yuhan and Ray, Siddhant and Cheng, Yihua and Zhang, Qizheng and Du, Kuntai and Lu, Shan and Jiang, Junchen},
  booktitle = {Proc. ACM EuroSys},
  pages={94--109},
  year={2025}
}

\end{document}